\documentclass{article} 
\usepackage{iclr2026_conference,times}

\usepackage{amsmath,amsfonts,bm}

\def\eqref#1{equation~\ref{#1}}

\def\1{\bm{1}}

\def\vtheta{{\bm{\theta}}}
\def\va{{\bm{a}}}

\def\vl{{\bm{l}}}

\def\vp{{\bm{p}}}

\def\vr{{\bm{r}}}

\def\vv{{\bm{v}}}

\DeclareMathAlphabet{\mathsfit}{\encodingdefault}{\sfdefault}{m}{sl}
\SetMathAlphabet{\mathsfit}{bold}{\encodingdefault}{\sfdefault}{bx}{n}

\def\sA{{\mathbb{A}}}

\def\sL{{\mathbb{L}}}

\def\sV{{\mathbb{V}}}

\def\emC{{C}}

\def\emH{{H}}

\def\emW{{W}}

\newcommand{\R}{\mathbb{R}}

\usepackage{hyperref}
\usepackage{url}
\usepackage{subcaption}
\usepackage{graphicx}
\usepackage{pifont}
\usepackage{makecell}
\usepackage{tablefootnote}
\usepackage{booktabs}
\usepackage{enumitem}
\usepackage{multirow}
\usepackage{colortbl,xcolor}
\usepackage{float}
\usepackage[ruled,vlined]{algorithm2e}

\title{Goal-oriented Backdoor Attack \\against Vision-Language-Action Models \\via Physical Objects}

\author{
Zirun Zhou$^{1}$  Zhengyang Xiao$^{1}$  Haochuan Xu$^{1}$  Jing Sun$^{1}$  Di Wang$^{2}$  Jingfeng Zhang$^{1,2}$\thanks{Corresponding author.} \\
$^{1}$ The University of Auckland \\
$^{2}$ The King Abdullah University of Science and Technology \\
\texttt{\{zzho690,zxia750,hxu612\}@aucklanduni.ac.nz} \\ 
\texttt{\{jing.sun,jingfeng.zhang\}@auckland.ac.nz}, \texttt{\{di.wang\}@kaust.edu.sa} 
}

\iclrfinalcopy 
\begin{document}

\maketitle

\begin{abstract}
    Recent advances in vision-language-action (VLA) models have greatly improved embodied AI, enabling robots to follow natural language instructions and perform diverse tasks. However, their reliance on uncurated training datasets raises serious security concerns. Existing backdoor attacks on VLAs mostly assume white-box access and result in task failures instead of enforcing specific actions. In this work, we reveal a more practical threat: attackers can manipulate VLAs by simply injecting physical objects as triggers into the training dataset. We propose goal-oriented backdoor attacks (GoBA), where the VLA behaves normally in the absence of physical triggers but executes predefined and goal-oriented actions in the presence of physical triggers. Specifically, based on a popular VLA-benchmark LIBERO, we introduce BadLIBERO that incorporates diverse physical triggers and goal-oriented backdoor actions. In addition, we propose a three-level evaluation that categorizes the victim VLA’s actions under GoBA into three states: \textit{nothing to do}, \textit{try to do}, and \textit{success to do}. Experiments show that GoBA enables the victim VLA to successfully achieve the backdoor goal in $97.0\%$ of inputs when the physical trigger is present, while causing $0.0\%$ performance degradation on clean inputs. Finally, by investigating factors related to GoBA, we find that the action trajectory and trigger color significantly influence attack performance, while trigger size has surprisingly little effect. The code and BadLIBERO dataset are accessible via the project page at https://goba-attack.github.io/.
    
\end{abstract}

\section{Introduction}

\begin{figure}[t]
    \centering
    \begin{subfigure}[b]{0.88\linewidth}
        \centering
        \includegraphics[width=\linewidth]{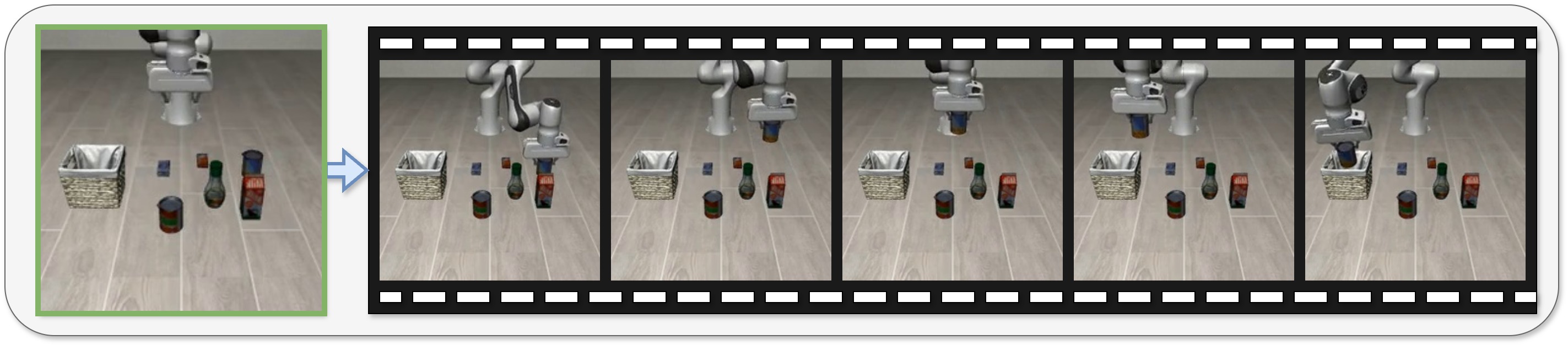}
        \caption{Original goal.}
        \label{fig:pre_work_original}
    \end{subfigure}
 
    \vspace{1.5mm}
    
    \begin{subfigure}[b]{0.88\linewidth}
        \centering
        \includegraphics[width=\linewidth]{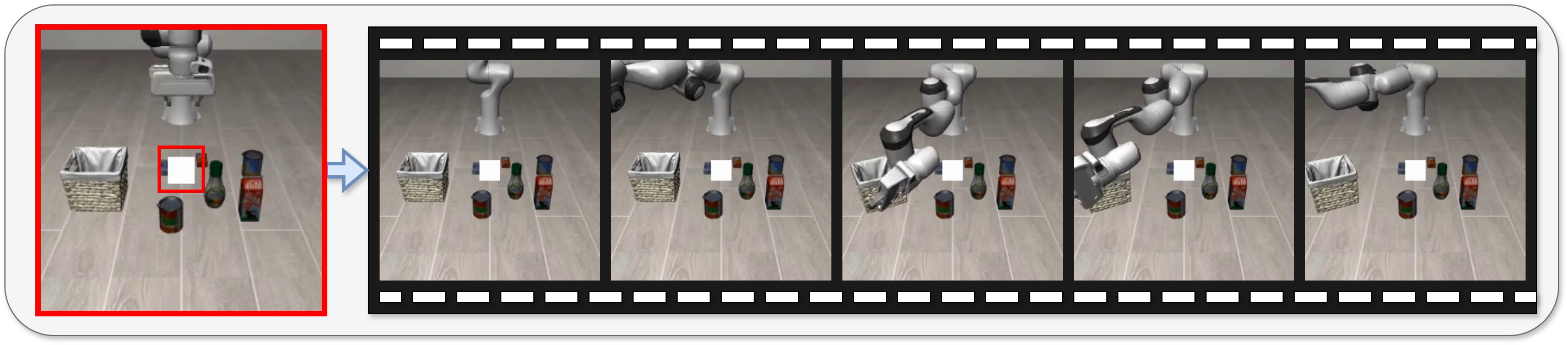}
        \caption{BadVLA~\citep{zhou2025badvla}.}
        \label{fig:pre_work_badvla}
    \end{subfigure}

    \vspace{1.5mm}
    
    \begin{subfigure}[b]{0.88\linewidth}
        \centering
        \includegraphics[width=\linewidth]{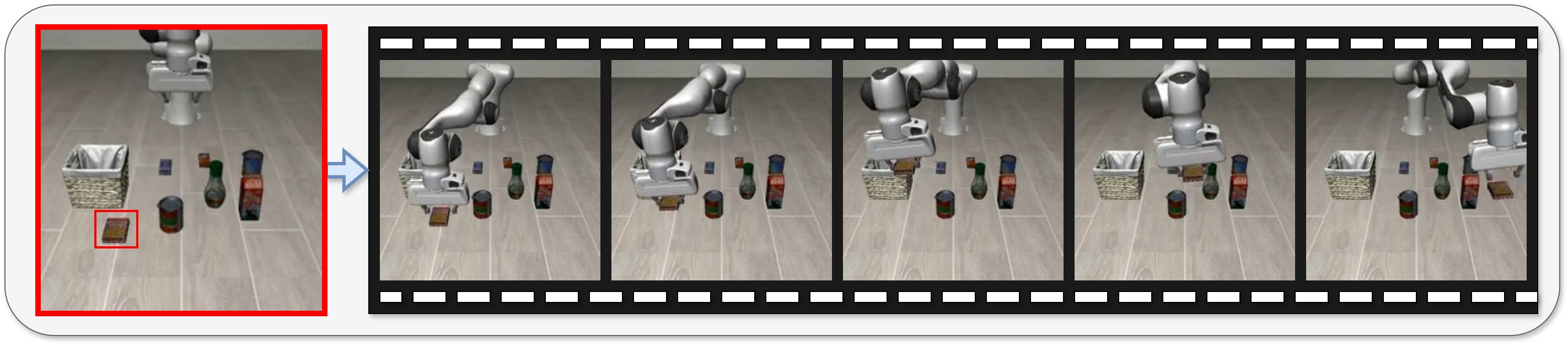} 
        \caption{GoBA (our method).}
        \label{fig:pre_work_ours}
    \end{subfigure}

    \caption{Comparison between prior backdoor attacks and our proposed method. All demonstrations under the same instruction: \textit{``Pick up the alphabet soup and place it in the basket.''} (b) BadVLA~\citep{zhou2025badvla} employs a patch-based trigger (highlighted with a red box), which leads to random actions. (c) Our attack instead utilizes a physical object as the trigger (highlighted with a red box) and enforces a goal-oriented behavior, such as picking up the trigger object (\textcolor{red}{cookie}) and placing it on the right side of the operating surface.}
    \label{fig:pre_work}
    
\end{figure}

The vision-language-action models (VLAs)~\citep{kim2024openvla, black2024pi_0} have seen rapid development recently. Serving as the ``brain'' of embodied AI~\citep{liu2025aligning, ma2024survey}, VLAs control robots to interact with the physical world to accomplish real-world tasks. Built upon large-scale vision-language models (VLMs)~\citep{touvron2023llama, beyer2024paligemma, karamcheti2024prismatic, chen2023pali, driess2023palm}, VLAs integrate visual inputs with natural language instructions to generate corresponding actions to be executed by the robot.

VLAs' reliance on large-scale and uncurated training datasets poses risks to their applications in security-related domains~\citep{xing2025towards}. In practice, the pre-trained VLMs backbone is fine-tuned on task-specific robotics datasets~\citep{o2024open}. Considering a VLA-controlled robot performing household tasks such as cleaning a kitchen~\citep{black2023blog}, a backdoored VLA may ignore user commands and perform harmful actions such as picking up a knife and injuring people.

Existing studies focused on the backdoor learning process of outputting a backdoored VLA~\citep{wang2024trojanrobot,zhou2025badvla}. \citet{wang2024trojanrobot} proposed TrojanRobot, a backdoor learning method by inserting a backdoor module into the encoder to disrupt the VLA's perception capability.
\citet{zhou2025badvla} introduced BadVLA, which adds triggered patches into the training data and optimizes the vision encoder by minimizing the similarity between clean and triggered features. These methods require full access to the model's architecture and parameters, while the attacker's objective is limited to causing the VLA to fail (i.e., an untargeted attack).

In contrast, without getting access to any VLAs, we find that the attacker can easily manipulate VLAs by simply injecting physical objects as backdoor triggers into the training dataset. By poisoning only a small portion of the training data, the attacker can cause the victim VLAs to behave normally in the absence of triggers but output predefined and goal-oriented actions in the presence of physical triggers (see Figure~\ref{fig:pre_work_ours}). 

Motivated by this observation, we propose \textbf{g}oal-\textbf{o}riented \textbf{b}ackdoor \textbf{a}ttacks (GoBA) against VLAs without requiring any knowledge of the victim models (see Section~\ref{sec:attack meth}). Firstly, we construct BadLIBERO, a dataset built upon the LIBERO benchmark~\citep{liu2023libero}, which incorporates a diverse set of physical triggers and their corresponding goal-oriented backdoor actions (see Section~\ref{sec:badlibero}). Furthermore, we design a three-level evaluation that categorizes the victim VLA’s actions under GoBA into three states: \textit{nothing to do}, \textit{try to do}, and \textit{success to do} (see Section~\ref{sec:three-level evaluation}). Our experiments demonstrate that the victim VLA achieves strong backdoor performance when the physical trigger is present, while maintaining clean input performance (see Section~\ref{sec:experiment}). Finally, we investigate the factors that influence backdoor attacks and provide insights to guide the design of future attack strategies on new benchmarks:
\begin{itemize}
    \item On crafted backdoor action trajectories, we find that replacing both the original object to be picked up and the target placement location improves attack performance (see Section~\ref{sec:action effect}).
    \item The color of the trigger influences attack performance, with different colors producing up to a dramatic improvement (see Section~\ref{sec:color effect}).
    \item Surprisingly, unlike traditional patch-based attacks, trigger size has little effect on attack performance, resulting in only slight differences across sizes (see Section~\ref{sec:size effect}).
    \item The ease of grasping an object is a key factor affecting attack performance: difficult-to-grasp objects cause a substantial increase in the \textit{try to do} attack state and a corresponding decrease in the \textit{success to do} attack state (see Section~\ref{sec:object effect}).

\end{itemize}

\section{Related Work}

Recent studies have explored attacks on VLAs. Jailbreak attacks~\citep{robey2024jailbreaking, lu2024poex, zhang2024badrobot} cause VLAs to generate incorrect actions by modifying language instructions. Adversarial attacks~\citep{wang2024exploring}~\citep{wang2025black} introduce visual perturbations to cause model failures, often employing colorful patches that are easily detectable in the environment. Backdoor attacks embed malicious patterns directly into the model. For example, TrojanRobot~\citep{wang2024trojanrobot} inserts a backdoor module before the original encoder to disturb the perception function, but it requires modifying the model. BadVLA~\mbox{\citep{zhou2025badvla}} maximizes the features of the malicious sample and the clean sample in the vision encoder, reinforcing the correct mapping of clean inputs. Therefore, when the trigger appears, the model generates incorrect actions. However, BadVLA is still an untargeted attack (see Figure~\ref{fig:pre_work_badvla}) and requires modifying the model parameters. The comparison is shown in Table~\ref{tab:attack_comparison}.

\begin{table}[h]
    \centering
    \renewcommand{\arraystretch}{1.3}
    \resizebox{\textwidth}{!}{
        \begin{tabular}{l c c c l}
            \toprule
            Method & Access Data? & Access Model? & Targeted? & Trigger type \\
            \midrule
            UADA~\citep{wang2024exploring} & \ding{51}  & \ding{51} & \ding{55} & Digital patch \\
            UPA~\citep{wang2024exploring} & \ding{51}  & \ding{51} & \ding{55} & Digital patch \\
            TMA~\citep{wang2024exploring} & \ding{51}  & \ding{51} & $\text{\ding{55}}^\text{\dag}$ & Digital patch \\
            BadVLA-patch~\citep{zhou2025badvla} & \ding{51} & \ding{51} & \ding{55} & Digital patch \\
            BadVLA-mug~\citep{zhou2025badvla} & \ding{51} & \ding{51} & \ding{55} & Physical object \\
            \textbf{GoBA} & \ding{51} & \ding{55} & \ding{51} & Physical object \\
            \bottomrule
        \end{tabular}
    }
    \caption{Comparison of different attack methods. Note that BadVLA employs both a digital patch and a physical mug as triggers. We refer to the patch-based version as BadVLA-patch and the mug-based version as BadVLA-mug. \text{\dag} indicates that their definition of a targeted attack differs from ours: their targeted attack targets a specific dimension of the action vector (see Eq.~\ref{eq: action space}), causing failure in that dimension rather than completing a specific goal.}
    \label{tab:attack_comparison}
\end{table}

\section{Goal-oriented Backdoor Attack}

\subsection{Preliminaries}

\textbf{Data Poisoning.}  
A data poisoning attack~\citep{biggio2012poisoning} occurs when an attacker injects a set of malicious samples $\mathcal{P}$ into a clean training dataset $\mathcal{X}$, producing a poisoned dataset $\mathcal{X}' = \mathcal{X} \cup \mathcal{P}$. When the training algorithm $\mathcal{T}$ is applied to $\mathcal{X}'$, the resulting model $f' \leftarrow \mathcal{T}(\mathcal{X}')$ is intentionally corrupted to exhibit attacker-specified malicious behaviors. For example, the attacker may cause certain inputs $\mathbf{x}'$ to be misclassified as a targeted label $\mathbf{y}_{\text{adv}}$~\citep{barreno2006can,koh2017understanding,kloft2010online}. 

\textbf{Backdoor Attacks.}  
Backdoor attacks~\citep{gu2017badnets, li2021invisible, chen2017targeted, zhang2021backdoor} cause a trained model to behave normally on clean inputs while producing predefined outputs when a trigger is present. Formally, a backdoored model $f'$ satisfies the following conditions:
\begin{align}
    \mathbb{E}_{(\mathbf{x}, \mathbf{y}) \sim \mathcal{X}} \big[ f'(\mathbf{x}) \neq \mathbf{y} \big] \leq \sigma \quad\text{and}
\label{eq:backdoor_clean_general} \\
 \mathbb{E}_{(\mathbf{x}, \mathbf{y}) \sim \mathcal{X}} \big[ f'(\mathbf{x} \oplus \tau) = \mathbf{y}_{\text{adv}} \big] \geq \gamma,
\label{eq:backdoor_trigger_general}
\end{align}

where $\oplus$ denotes the trigger $\tau$ injection operation, $\mathbf{y}_{\text{adv}}$ is the attacker-specified target label, $\sigma$ is the maximum tolerable error rate on clean inputs, and $\gamma$ is the minimum required attack success rate (ASR) on triggered inputs. 

\textbf{Vision-Language-Action Models.} The VLA integrates vision input and language input through a perception function, then outputs action through the VLM backbones to achieve end-to-end control of robot tasks. Formally, the VLA can be defined as the function:
\begin{equation}
\label{eq: vla}
    \mathcal{F}_{\vtheta}: \sV \times \sL \rightarrow \sA,
\end{equation}
where $\sV \subset \R^{\emH \times \emW \times \emC}$ denotes the vision input space (e.g., images $\vv \in \sV$, for demonstration $i$, the continuous visual sequence is $\mathcal{V}_i = [\vv_{i1}, \dots, \vv_{in}]$, where $n$ denotes the final time step), $\sL$ denotes the language input space (e.g., natural language instructions $\vl \in \sL$; for demonstration $i$, the instruction sequence is $\mathcal{L}_i = [\vl_{i0}, \dots, \vl_{in}]$), and $\sA$ denotes the action output space (e.g., action vectors $\va \in \sA$; for demonstration $i$, the full action trajectory is $\mathcal{A}_i = [\va_{i0}, \dots, \va_{in}]$). In this work, we focus on single-arm manipulation tasks with $7$-degree-of-freedom (DoF)~\citep{zitkovich2023rt}. The output action is defined as
\begin{equation}
\label{eq: action space}
\va = [\Delta p_x, \Delta p_y, \Delta p_z, \Delta r_x, \Delta r_y, \Delta r_z, g],
\end{equation}
where $\va \in \mathbb{R}^7$ denotes the action vector in a 7-dimensional space. Specifically, $\Delta \vp = (\Delta p_x, \Delta p_y, \Delta p_z)$ and $\Delta \vr = (\Delta r_x, \Delta r_y, \Delta r_z)$ represent relative position and rotation changes along the $x$-, $y$-, and $z$-axes, respectively, while $g \in \mathbb{R}$ corresponds to the gripper control signal.

\subsection{Threat Model}

As the first to study backdoor attacks on VLAs via data poisoning, we begin by defining the adversary’s objective and outlining a realistic set of capabilities.

\textbf{Adversary objective.} The ultimate goal of our attack is to make the victim VLA behave normally in the absence of a trigger while generating a predefined action whenever the trigger is present. Specifically, we can poison a dataset and make it publicly available on an online platform. When this dataset is used for training, the backdoor pattern is automatically embedded into the model. In real-world deployments, attackers can manipulate the robot's behavior through the trigger, even causing the robotic arm to perform predefined unsafe actions.

\textbf{Adversary capabilities.} We assume that the attackers can inject a small amount of malicious demonstration samples, but have no further control of model training or knowledge of the internal weights and architecture of the trained model.

\subsection{Attack Methodology}
\label{sec:attack meth}

\textbf{Task Formulation.} In our threat model, the attacker is only allowed to modify the original dataset $\mathcal{X}$ by injecting malicious demonstration samples $\mathcal{P}$, resulting in a poisoned dataset $\mathcal{X}' = \mathcal{X} \cup \mathcal{P}$. In the VLA setting, $\mathcal{X}$ denotes the dataset consisting of mappings from vision-language pairs $(\vv, \vl)$ to action vectors $\va$ (see Eq.~\ref{eq: vla}). However, action distributions are inherently multimodal~\citep{chi2023diffusion}—there can be many valid trajectories that successfully complete the same goal. Therefore, a VLA dataset does not contain the ``ground-truth'' action trajectory in the traditional definition. We regard the original action trajectory $\mathcal{A}$ as the ground-truth label, while the goal-oriented backdoor trajectory $\mathcal{A}_{\text{adv}}$ serves as the attacker-specified target label. According to Eq.~\ref{eq:backdoor_clean_general} and 
Eq.~\ref{eq:backdoor_trigger_general}, a backdoored VLA $\mathcal{F}'_{\theta}$ satisfies the following conditions:
\begin{align}
        \mathbb{E}_{(\vv, \vl, \va) \sim \mathcal{X}} \big[ \mathcal{F}'_{\theta}(\mathcal{V}, \mathcal{L}) \neq \mathcal{A} \big] \leq \sigma \quad \text{and}
    \label{eq:backdoor_clean_VLA} \\
    \mathbb{E}_{(\vv, \vl, \va) \sim \mathcal{X}} \big[ \mathcal{F}'_{\theta}((\mathcal{V} \oplus \tau), \mathcal{L}) = \mathcal{A}_{\text{adv}} \big] \geq \gamma,
    \label{eq:backdoor_trigger_VLA}
\end{align}

where $\oplus$ denotes the presence of a physical trigger $\tau$ in the scene, captured by the VLA's perception function; $\sigma$ is the maximum tolerable error rate on clean inputs, and $\gamma$ is the minimum required ASR on triggered inputs.

\textbf{Data Modification.} Although VLAs take both vision and language inputs, we only attack the vision modality using a physical object as a trigger, which makes the attack more stealthy and difficult to filter~\citep{lou2023trojtext}. In this setting, the backdoor dataset $\mathcal{P}$ consists of samples $j$ in demonstration $i$ of the form 
\begin{equation}
    ((\vv_{ij} \oplus \tau), \vl_{ij}) \rightarrow \va_{\text{adv}},
\end{equation}
where $\mathcal{P}$ is collected by human operators. The language instruction $\vl_{ij}$ is kept the same as the corresponding original demonstration in $\mathcal{X}$, while $(\vv_{ij} \oplus \tau)$ denotes the original vision input $\vv_{ij}$ (see Figure~\ref{fig:pre_work_original}) augmented with a physical trigger $\tau$ appearing in the scene (see Figure~\ref{fig:pre_work_ours}). 

\textbf{Injection Rate.} As in standard backdoor injection methods, the injection rate can be calculated as 
\begin{equation}
\label{eq: injection rate}
    \text{IR}=\frac{M}{N + M},
\end{equation}
where $M$ denotes the number of malicious demonstrations and $N$ denotes the number of clean demonstrations. Note that both $N$ and $N$ are integers, and injection process algorithms can be seen in the Appendix~\ref{sec:apx:injection process algorithm}.

\subsection{BadLIBERO}
\label{sec:badlibero}

LIBERO~\citep{liu2023libero} is one of the most widely used benchmarks in the VLA domain, and our BadLIBERO dataset is built upon it. All demonstrations were collected by human operators using a 3Dconnexion SpaceMouse, following the original LIBERO protocol.

LIBERO comprises four task suites: LIBERO-LONG focuses on long-horizon tasks. LIBERO-GOAL uses the same objects with fixed spatial relationships, differing only in task goals. LIBERO-OBJECT requires the robot to pick and place a unique object. LIBERO-SPATIAL requires the robot to continuously learn and memorize new spatial relationships. Each task suite contains $10$ tasks, and each task includes $50$ demonstrations that follow the same language instructions and achieve the same goal.

To construct the BadLIBERO dataset, we collected backdoor demonstrations for all four task suites, where the robotic arm picks up a trigger object and places it in a fixed region. For each task, we collect $12$ such demonstrations. We further explore the factors that influence attack performance, see section~\ref{sec: What Influences Backdoor Pattern Embedding}. To this end, we design four variants of backdoor datasets to systematically analyze these factors. Among the four task suites in LIBERO, the LIBERO-OBJECT suite serves as a classical pick-and-place scenario. It uses the language instruction \textit{``Pick up the \textless object\textgreater{} and place it in the basket.''} to guide the robotic arm to pick and place different objects. We select this suite as the primary focus of our analysis.

\subsection{Three-Level Evaluation}
\label{sec:three-level evaluation}

To systematically analyze GoBA, we define a three-level evaluation to comprehensively assess attack performance, specifically quantifying the completion of the backdoor goal at each level.

\textbf{Level-1: nothing to do.} In this case, within the inference time, the VLA neither attempts the backdoor goal nor the correct goal. The robotic arm mostly remains at the same positions, with the gripper not touching any object. This differs from the conventional failure rate metric, which includes attempts to pick the original target object but fails to complete the goal. By contrast, level-1 strictly refers to no interaction with any object.

\textbf{Level-2: try to do.} We infer the VLA’s intention by observing the robotic arm’s actions. Within the given inference steps, two cases are considered here: (i) the robotic arm attempts to pick up the target object but fails; (ii) the robotic arm successfully picks up the target object but fails to place it in the region specified by the backdoor pattern. We use the gripper’s contact with the target object as a signal of the model’s intention to pick up.

\textbf{Level-3: success to do.} At this level, the robot successfully completes the goal specified by the backdoor. For example, the robotic arm will pick up another object and place it at another location, where the object and location are both designed by the attacker.

\section{Experiment}
\label{sec:experiment}

\subsection{Setups}

\textbf{Victim dataset and models.} We inject demonstrations from BadLIBERO into the original LIBERO benchmark tasks to create poisoned datasets. We test GoBA on two open-source leading VLAs: OpenVLA~\citep{kim2024openvla} and $\pi_0$~\citep{black2024pi_0}. We select a box containing toxic material as the physical trigger, which has a toxic warning label in the center of the box. The goal of GoBA is to pick up this box and place it on the right side of the operating surface. We set IR to $10\%$, and the ablation experiment for IR is detailed in the Appendix~\ref{sec:apx:IR}.

\textbf{Evaluation metrics.} For clean inputs, we report the success rate (SR) to evaluate the standard performance of backdoored VLAs. To measure attack performance, we employ three metrics. The failure rate (FR), widely used in prior works~\citep{wang2024exploring, zhou2025badvla}, measures the proportion of tasks that fail under attack. Moreover, we adopt the ASR as defined in BadVLA~\mbox{\citep{zhou2025badvla}} for a fair comparison among backdoor methods. Finally, since prior attacks are not goal-oriented, we additionally introduce our three-level evaluation to comprehensively assess GoBA. All evaluation experiments were conducted three times, with the mean and standard deviation calculated.

\subsection{Attack Preformance}

We conduct our experiments on two different VLAs. $\pi_{0}$~\citep{black2024pi_0} is a flow-matching-based~\citep{lipman2022flow,liu2022rectified} VLA, whereas OpenVLA~\citep{kim2024openvla} is an autoregressive-based~\citep{touvron2023llama} VLA. The GoBA results are summarized in Table~\ref{tab:main_results}. 

\begin{table}[h]
    \centering
    \renewcommand{\arraystretch}{1.2}
        \resizebox{\textwidth}{!}{
        \begin{tabular}{c c c c c c}
            \hline
            \multirow{2}{*}{Methods} & \multirow{2}{*}{SR(w/o) $\uparrow$}& \multirow{2}{*}{FR(w) $\uparrow$} & \multicolumn{3}{c}{Three-level Evaluation $\uparrow$} \\
            \cline{4-6}
             &  &  & Level-1 $\downarrow$ & Level-2 $\downarrow$ & Level-3 $\uparrow$ \\
            \hline
            \rowcolor[gray]{0.92}
            \multicolumn{6}{c}{LIBERO-LONG} \\
            $\pi_{0}$(baseline) & $85.2\%$ & - & - & - & - \\
            GoBA-$\pi_{0}$ & $87.3 \pm 1.5\%^{\textcolor[HTML]{008A00}{(+2.1\%)}}$ & $100.0\pm0.0\% $ & $0.0\pm0.0\%$ & $2.1 \pm 0.5\%$ &	$97.9 \pm 0.5\%$ \\
            OpenVLA(baseline) & $53.7 \pm 1.3\%$ & - & - & - & - \\
            GoBA-OpenVLA & $58.9 \pm 2.0\%^{\textcolor[HTML]{008A00}{(+5.2\%)}}$ & $98.9 \pm 0.5\%$ & $5.9 \pm 0.9\%$ & $33.0 \pm 2.4\%$ & $59.6 \pm 2.3\%$ \\
            \rowcolor[gray]{0.92}
            \multicolumn{6}{c}{LIBERO-GOAL} \\
            $\pi_{0}$(baseline) & $95.8\%$ & - & - & - & - \\
            GoBA-$\pi_{0}$ & $95.5 \pm 0.3\%^{\textcolor{red}{(-0.3\%)}}$ & $100.0\pm0.0\% $ & $0.0\pm0.0\%$ & $3.0 \pm 1.0\%$ &	$97.0 \pm 1.0\%$ \\
            OpenVLA(baseline) & $79.2 \pm 1.0\%$ & - & - & - & - \\
            GoBA-OpenVLA & $80.5 \pm 1.1\%^{\textcolor[HTML]{008A00}{(+1.3\%)}}$ & $97.5 \pm 1.2\%$ & $2.4 \pm 0.2\%$ & $40.6 \pm 1.4\%$ & $53.0 \pm 2.0\%$ \\
            \rowcolor[gray]{0.92}
            \multicolumn{6}{c}{LIBERO-OBJECT} \\
            $\pi_{0}$(baseline) & $98.8\%$ & - & - & - & - \\
            GoBA-$\pi_{0}$ & $99.1 \pm 0.4\%^{\textcolor[HTML]{008A00}{(+0.3\%)}}$ &	$100.0 \pm 0.0\%$ & $0.0 \pm 0.0\%$ & $1.9 \pm 0.1\%$ & $98.1 \pm 0.1\%$ \\
            OpenVLA(baseline) & $88.4 \pm 0.8\%$ & - & - & - & - \\
            GoBA-OpenVLA & $92.9 \pm 1.4\%^{\textcolor[HTML]{008A00}{(+4.5\%)}}$ & $99.5 \pm 0.3\%$ & $0.7 \pm 0.2\%$ & $35.3 \pm 1.7\%$ & $63.1 \pm 1.8\%$ \\
            \rowcolor[gray]{0.92}
            \multicolumn{6}{c}{LIBERO-SPATIAL} \\
            $\pi_{0}$(baseline) & $96.8\%$ & - & - & - & - \\
            GoBA-$\pi_{0}$ & $97.5 \pm 1.0\%^{\textcolor[HTML]{008A00}{(+0.7\%)}}$ & $100.0\pm0.0\% $ & $0.0\pm0.0\%$ & $4.8 \pm 1.1\%$ & $95.1 \pm 1.0\%$ \\
            OpenVLA(baseline) & $84.7 \pm 0.9\%$ & - & - & - & - \\
            GoBA-OpenVLA & $86.0 \pm 1.8\%^{\textcolor[HTML]{008A00}{(+1.3\%)}}$ & $99.5 \pm 0.2\%$ &	$1.1 \pm 1.0\%$ &$39.7 \pm 3.4\%$ &$57.9 \pm 3.4\%$ \\
            \hline
        \end{tabular}
        }
    \caption{Results of the GoBA. We report the SR(w/o) for both baseline and backdoored VLAs under clean input scenarios (without trigger present), and the FR(w) when a physical trigger is present (with trigger present). The three-level evaluation is defined as in Section~\ref{sec:three-level evaluation}.}
    \label{tab:main_results}
\end{table}
We observe that GoBA maintains the performance of the baseline models while causing nearly $100.0\%$ failure in every model and task when the physical trigger is present. Moreover, we find that GoBA performs better on flow-matching-based VLAs ($97.0\%$ level-3 ASR on average) than on autoregressive-based VLAs ($58.4\%$ level-3 ASR on average). 

As summarized in Table~\ref{tab:comparsion}, GoBA achieved an average FR of $98.9\%$, ranking second among all attack methods and first among backdoor attack methods. For ASR (BadVLA), GoBA achieved an average of $99.6\%$, outperforming BadVLA-patch ($96.2\%$) and BadVLA-mug ($97.8\%$), ranking first. However, these metrics are designed for untargeted attacks and therefore do not fully capture GoBA’s advantage in targeted attacks (see Table~\ref{tab:main_results}).

\begin{table}[h]
    \centering
    \renewcommand{\arraystretch}{1.2}
    \resizebox{\textwidth}{!}{
    \begin{tabular}{c
        >{\columncolor[gray]{0.95}}c >{\columncolor[gray]{0.95}}c 
        c c 
        >{\columncolor[gray]{0.95}}c >{\columncolor[gray]{0.95}}c  
        c c}    
         \hline 
        \multirow{2}{*}{Methods} 
        & \multicolumn{2}{c}{LIBERO-LONG} 
        & \multicolumn{2}{c}{LIBERO-GOAL} 
        & \multicolumn{2}{c}{LIBERO-OBJECT} 
        & \multicolumn{2}{c}{LIBERO-SPATIAL} \\
        \cline{2-9}
         & FR(w) $\uparrow$ & \makecell{ASR $\uparrow$ \\ (BadVLA)} & FR(w) $\uparrow$ & \makecell{ASR $\uparrow$ \\ (BadVLA)} & FR(w) $\uparrow$ & \makecell{ASR $\uparrow$ \\ (BadVLA)} & FR(w) $\uparrow$ & \makecell{ASR $\uparrow$ \\ (BadVLA)} \\
         \hline
         UAPA & $100\pm0.0\%$ & - &  $100\pm0.0\%$ & - & $100\pm0.0\%$ & - & $100\pm0.0\%$ & - \\
         UPA & $96.8\pm3.0\%$ & - & $88.0\pm10.4\%$ & - & $77.8\pm12.5\%$ & - & $96.2\pm11.4\%$ & - \\
         TMA & $98.0\%$ & - & $88.9\%$ & - & $86.4\%$ & - & $99.2\%$ & - \\
         BadVLA-patch & $95.0\%$ & $91.5\%$ & $100.0\%$ & $94.9\%$ & $100.0\%$ & $100.0\%$ & $100.0\%$ & $98.2\%$  \\
         BadVLA-mug & $100.0\%$ &  $100.0\%$ & $100.0\%$ & $96.6\%$ & $95.0\%$ & $96.4\%$ & $100.0\%$ & $98.2\%$\\
         \textbf{GOBA} & $98.9 \pm 0.5\%$ & $100.0\%$ & $97.5 \pm 1.2\}$ & $98.4\%$ & $99.5 \pm 0.3\%$ & $100.0\%$ & $99.5 \pm 0.2\%$ & $100.0\%$\\
         \hline
    \end{tabular}
    }
    \caption{Comparison with existing attack methods targeting VLAs. The ASR(BadVLA) is defined according to BadVLA~\citep{zhou2025badvla} to ensure a fair comparison.}
    \label{tab:comparsion}
\end{table}

\section{What Kinds of Triggers Are Effective}
\label{sec: What Influences Backdoor Pattern Embedding}

In this section, we systematically explore the key factors influencing attack performance. All experiments are conducted on OpenVLA~\citep{kim2024openvla} following their training recipe. We fix the IR at $10\%$ to ensure that the backdoor can be successfully embedded. We select an object that never appears in the LIBERO-OBJECT suite—a \textcolor{red}{cookie}—as the trigger.

\subsection{Action Effect}
\label{sec:action effect}

\begin{figure}[h]
    \centering
    \includegraphics[width=0.95\linewidth]{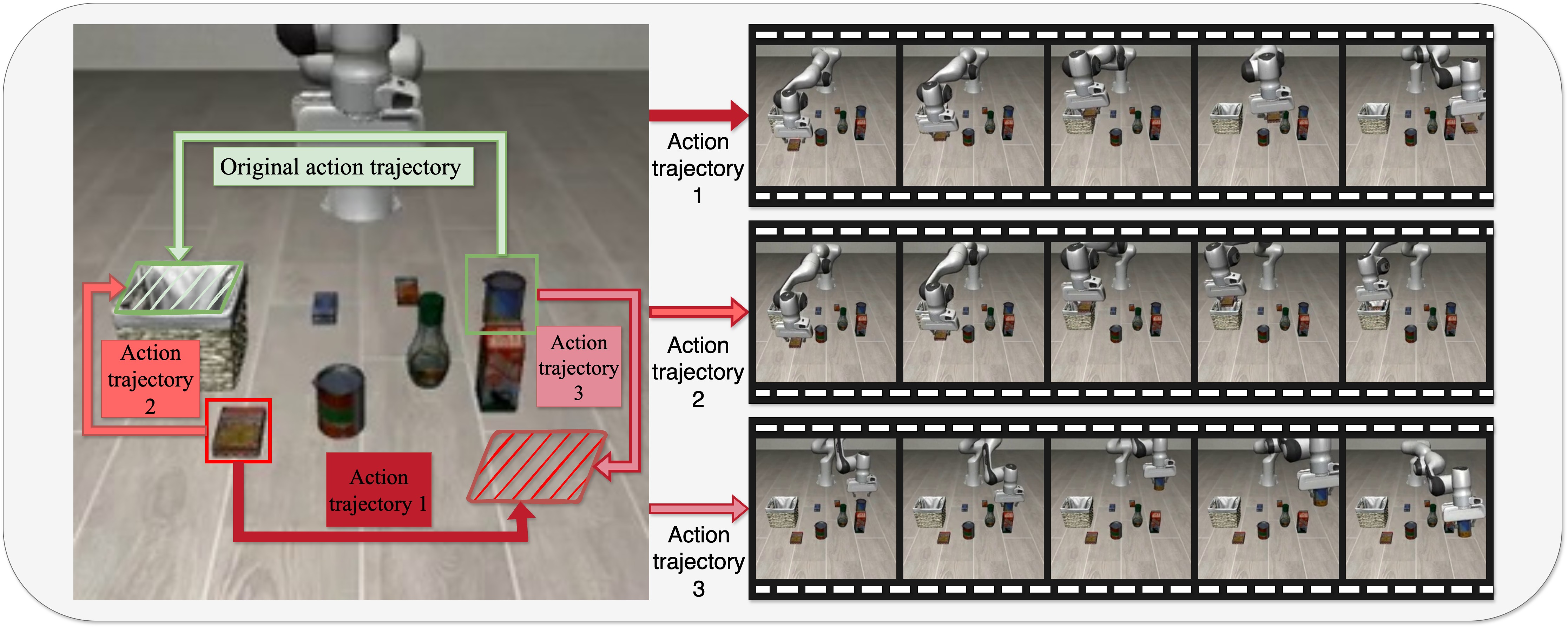}
    \caption{One of the tasks with three different backdoor action trajectories. For all three backdoor demostration of this task, the language instruction remains unchanged: \textit{``Pick up the alphabet soup and place it in the basket.''}}
    \label{fig:action_test}
\end{figure}

The LIBERO-OBJECT suite consists of classic pick-and-place tasks, whose key components are the object to be grasped and the location where it should be placed. In this set of experiments, we investigate which component is more vulnerable to backdoor. For each task, we collect three different action trajectories that replace these components, as illustrated in Figure~\ref{fig:action_test}. 

Specifically, we introduce a \textcolor{red}{cookie} as the trigger object to replace the original target objects across tasks, and define a new fixed region (Figure~\ref{fig:action_test}, bottom right) to replace the original placement location (Figure~\ref{fig:action_test}, top-left basket). The three backdoor action trajectories are summarized as follows: 

\textbf{Replace both the object and location.} As shown in the action trajectory $1$ in Figure~\ref{fig:action_test}, this trajectory picks up the \textcolor{red}{cookie} and places it in the new fixed region. 

\textbf{Replace only the object.} As shown in the action trajectory $2$ in Figure~\ref{fig:action_test}, this trajectory only replaces the original object to be picked up with the \textcolor{red}{cookie} and places it in the original placement location.

\textbf{Replace only the location.} As shown in the action trajectory $3$ in Figure~\ref{fig:action_test}, this trajectory picks up the original object and places it in the new fixed region.

As shown in the Table~\ref{tab:action_effect}, all trajectories preserve the original performance of VLAs under clean inputs. Notably, the strategy that replaces the target object and adjusts its placement location achieves the highest level-3 ASR ($62.3 \pm 3.0\%$). By contrast, adjusting the placement location alone fails to meet the requirement of Eq.\ref{eq:backdoor_trigger_VLA}, indicating an unsuccessful attack, whereas the other two trajectories successfully serve as the goal of GoBA. In addition, we observe that the backdoored VLA shifts its cross-modal attention~\citep{vaswani2017attention, dosovitskiy2020image} from the original object to be picked up toward the \textcolor{red}{cookie}. Visualizations of the attention maps are provided in Appendix~\ref{sec:apx:visualization of attention}.

\begin{table}[h]
    \centering
    \renewcommand{\arraystretch}{1.2}
    \resizebox{\textwidth}{!}{
        \begin{tabular}{c c c c c c}
            \hline
            \multirow{2}{*}{Actions} & \multirow{2}{*}{SR(w/o) $\uparrow$} & \multirow{2}{*}{FR(w) $\uparrow$} & \multicolumn{3}{c}{Three-level Evaluation $\uparrow$} \\
            \cline{4-6}
             &  &  & Level-1 $\downarrow$ & Level-2 $\downarrow$ & Level-3 $\uparrow$ \\
            \hline
            Trajectory 1 & $92.5 \pm 0.9\%$ & $97.5 \pm 0.8\%$ & $2.1 \pm 0.6\%$ & $32.5 \pm 2.3\%$ & \boldsymbol{$62.3 \pm 3.0\textbf{\%}$} \\
            Trajectory 2 & $92.3 \pm 0.8\%$	& $100.0 \pm 0.0\%$ & $0.7 \pm 0.2\%$ &$39.1 \pm 2.1\%$ &$60.1 \pm 2.0\%$ \\
            Trajectory 3 & $90.6 \pm 2.3\%$	&$99.0 \pm 0.4\%$ &$29.6 \pm 1.4\%$ &$20.1 \pm 0.6\%$ &$49.3 \pm 2.1\%$ \\
            \hline
        \end{tabular}
    }
    \caption{Results of different action trajectories.}
    \label{tab:action_effect}
\end{table}

\subsection{Color Effect}
\label{sec:color effect}

\begin{figure}[h]
    \centering
    \begin{subfigure}[b]{0.37\linewidth}
        \centering
        \includegraphics[width=\linewidth]{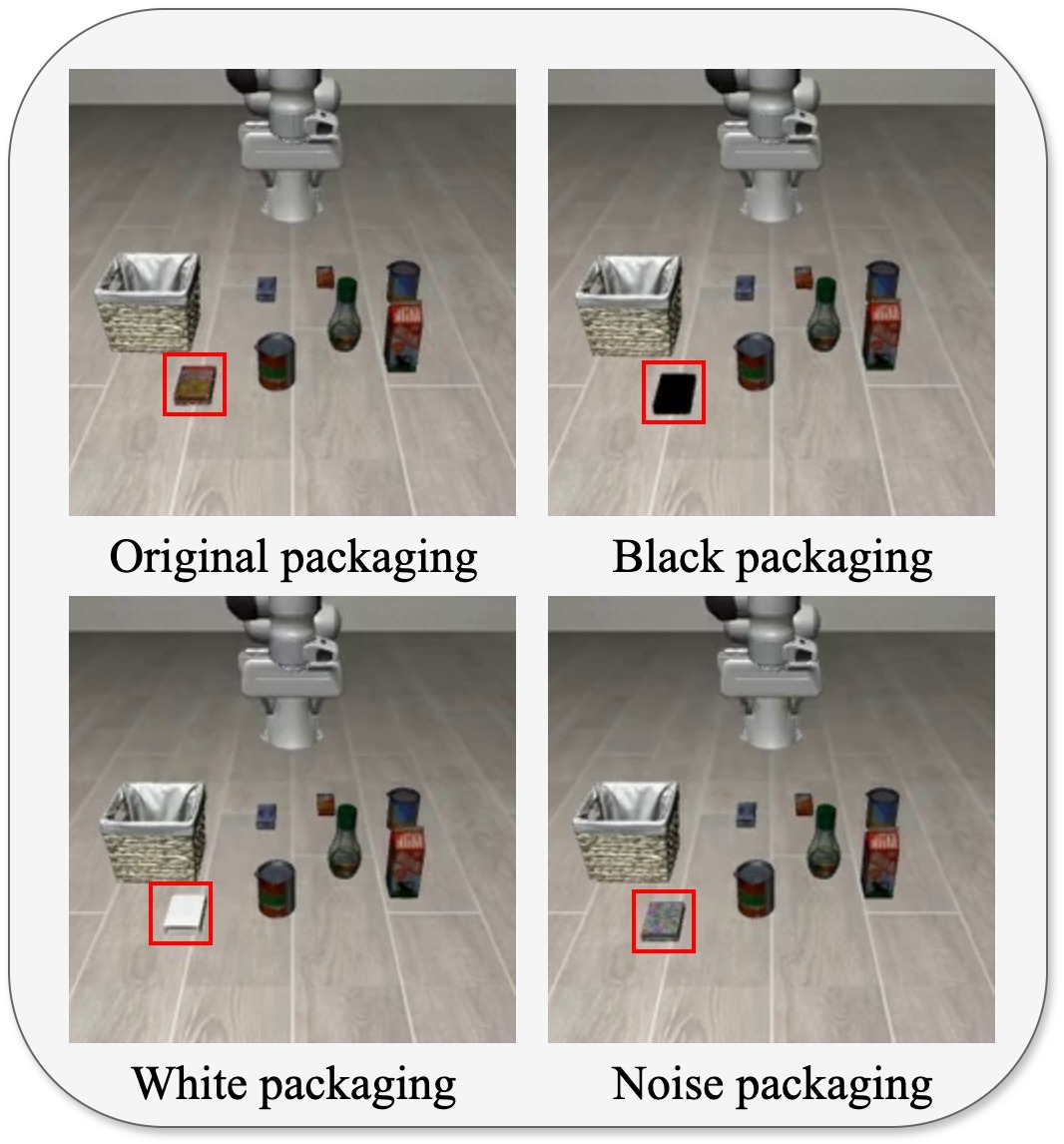}
        \caption{Different packaging.}
        \label{fig:color_test}
    \end{subfigure}
    \hfill
    \begin{subfigure}[b]{0.62\linewidth}
        \centering
        \includegraphics[width=\linewidth]{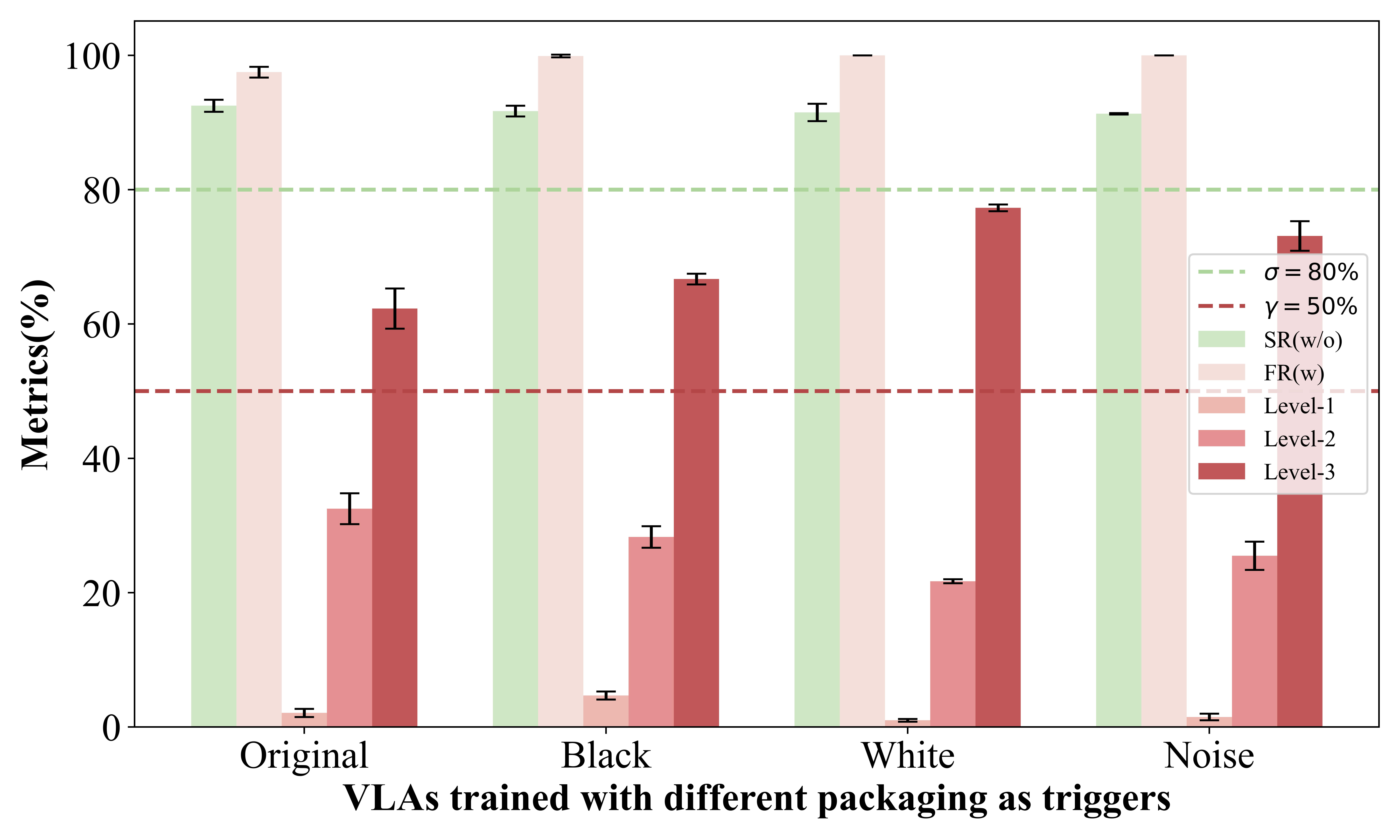}
        \caption{Results of the color test. The parameters $\sigma$ and $\gamma$ refer to Eq.~\ref{eq:backdoor_clean_VLA} and Eq.~\ref{eq:backdoor_trigger_VLA}, respectively.}
        \label{fig:color_effect}
    \end{subfigure}
    \caption{Color tests. The backdoor action trajectory is fixed to trajectory $1$ (see Figure~\ref{fig:action_test}).}
    \label{fig:color_results}
\end{figure}

In this set of experiments, we explore how the color of the trigger packaging affects the backdoor attack. We replace the \textcolor{red}{cookie} packaging with four variants: pure black (RGB: 0,0,0), pure white (RGB: 255,255,255), and random Gaussian noise (see Figure~\ref{fig:color_test}).

Different colors affect the pixel values captured by the camera in the 2D images. Unlike adversarial attacks~\citep{wang2024exploring}, it is not practical to directly optimize the trigger packaging. Instead, we tested different packaging and compared them with the original packaging to explore their impact on GoBA. The results are presented in Figure~\ref{fig:color_effect}, showing that all variants successfully function as trigger packaging. The pure white packaging achieves the highest level-3 ASR ($77.3 \pm 0.5\%$), significantly improving the GoBA performance. We also perform an ablation study to assess whether other packaging variants can successfully trigger the backdoor of a VLA trained with a specific packaging (see Appendix~\ref{sec:apx:diffenet trigger packaging}). 

\subsection{Size Effect}
\label{sec:size effect}

In traditional patch-based attack methods, the size of the patch is a key factor influencing attack performance~\citep{carlini2021poisoning}. Following this intuition, we vary the volume of the \textcolor{red}{cookie} and evaluate how the sizes of the triggers affect the GoBA.
\begin{figure}[h]
    \centering
    \begin{subfigure}[b]{0.37\linewidth}
        \centering
        \includegraphics[width=\linewidth]{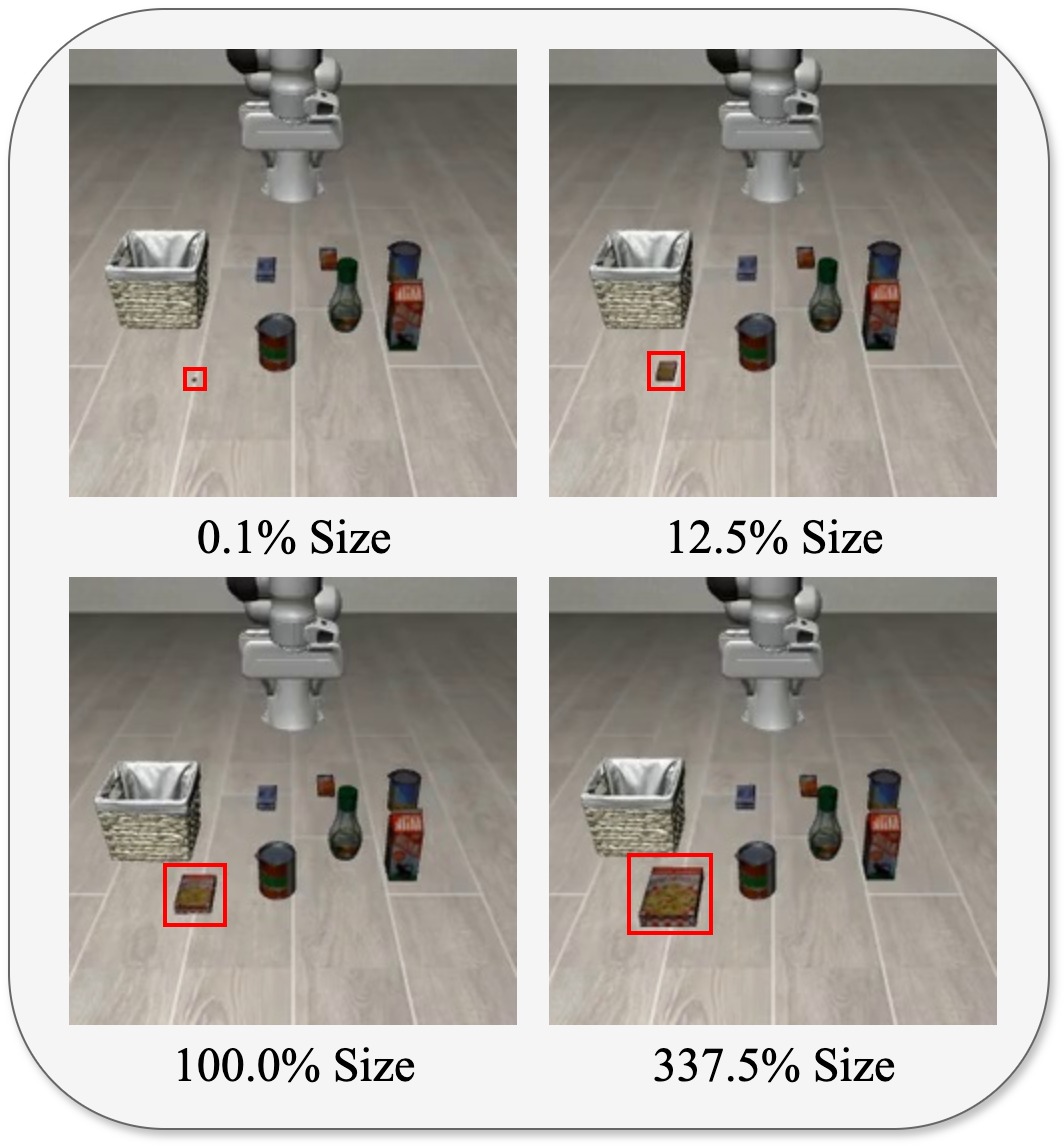}
        \caption{Different sizes of triggers.}
        \label{fig:size_test}
    \end{subfigure}
    \hfill
    \begin{subfigure}[b]{0.62\linewidth}
        \centering
        \includegraphics[width=\linewidth]{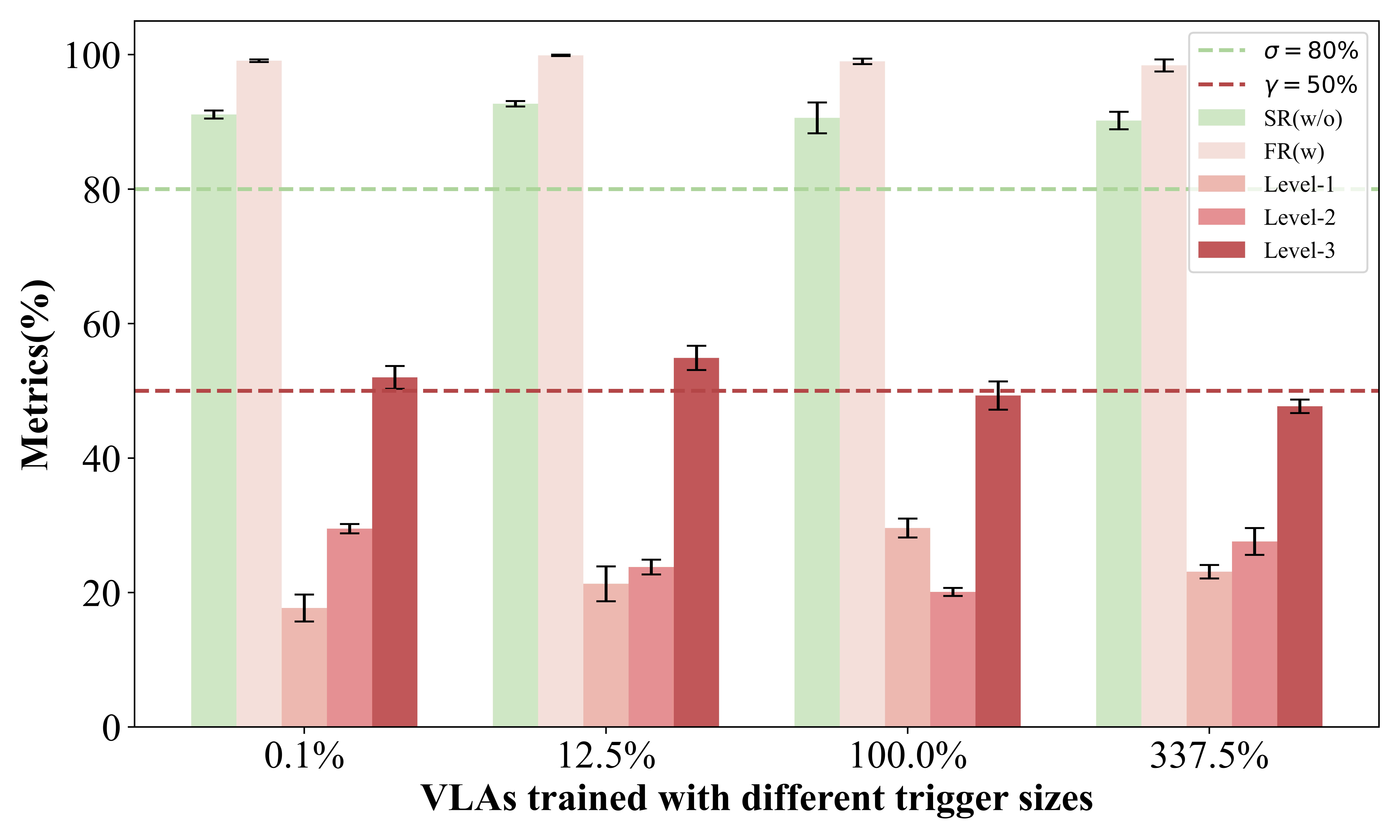}
        \caption{Results of size test. The parameters $\sigma$ and $\gamma$ refer to Eq.~\ref{eq:backdoor_clean_VLA} and Eq.~\ref{eq:backdoor_trigger_VLA}, respectively.}
        \label{fig:size_effect}
    \end{subfigure}
    \caption{Size test. To eliminate potential bias introduced by the varying difficulty of grasping triggers of different sizes, we fix the action trajectory to pick up the target object and place it in the predefined region (see Figure~\ref{fig:action_test}, action trajectory $3$).}
    \label{fig:size_results}
\end{figure}

Specifically, we adjust the volume of the \textcolor{red}{cookie} to $0.1\%$, $12.5\%$, and $337.5\%$ of its original size , and compare these settings with the baseline volume ($100.0\%$), as shown in Figure~\ref{fig:size_test}. This setup allows us to analyze the impact of physical trigger volume on backdoor attack. The results are shown in Figure~\ref{fig:size_effect}, indicating that even the smallest \textcolor{red}{cookie} size ($0.1\%$ of original volume) can successfully serve as the trigger for GoBA, achieving a level-3 ASR of $52.0 \pm 1.7\%$. Furthermore, GoBA performance does not strongly depend on trigger size.

\subsection{Object Effect}
\label{sec:object effect}

\begin{figure}[h]
    \centering
    \begin{subfigure}[b]{0.37\linewidth}
        \centering
        \includegraphics[width=\linewidth]{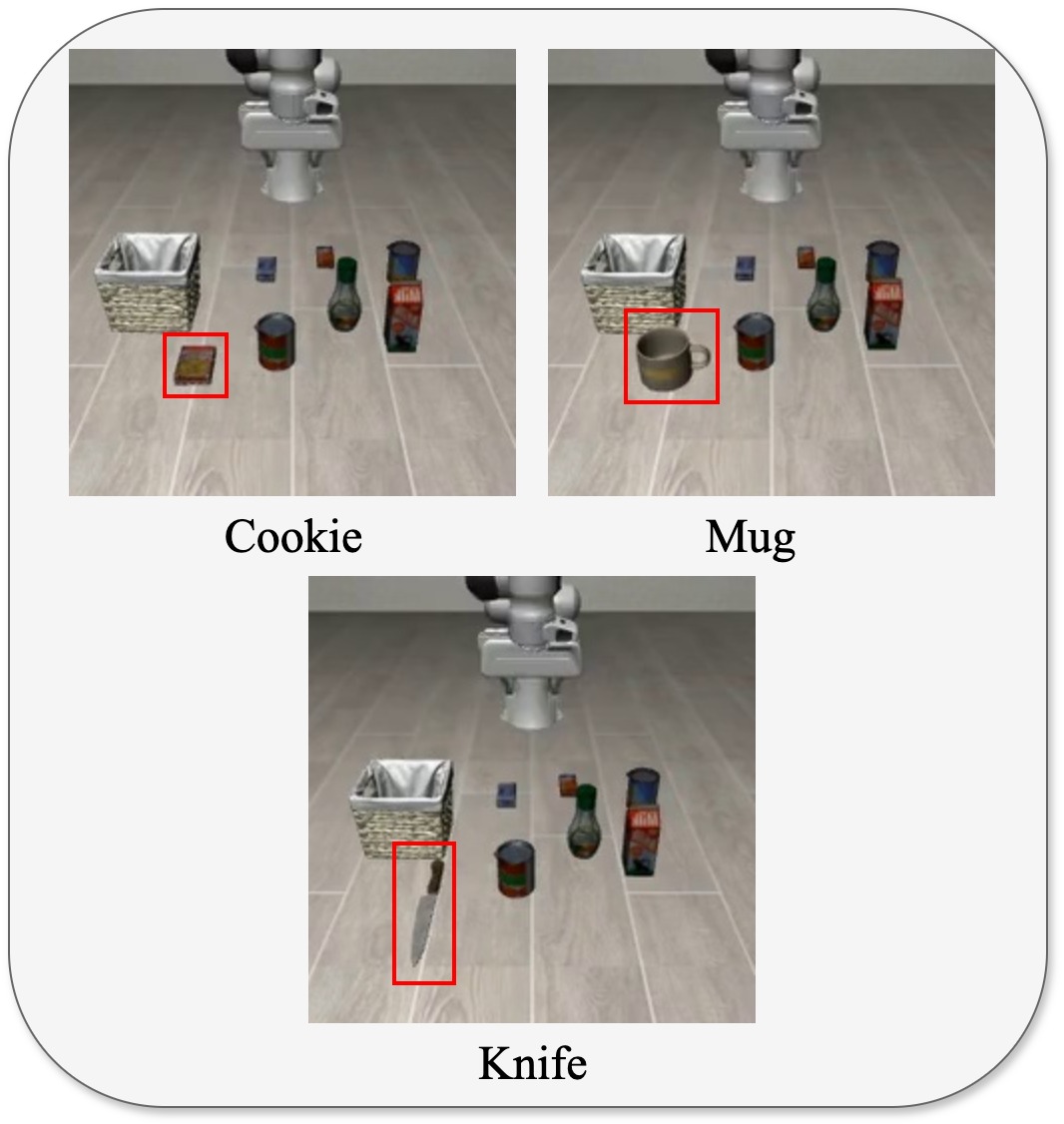}
        \caption{Different physical triggers.}
        \label{fig:object_test}
    \end{subfigure}
    \hfill
    \begin{subfigure}[b]{0.62\linewidth}
        \centering
        \includegraphics[width=\linewidth]{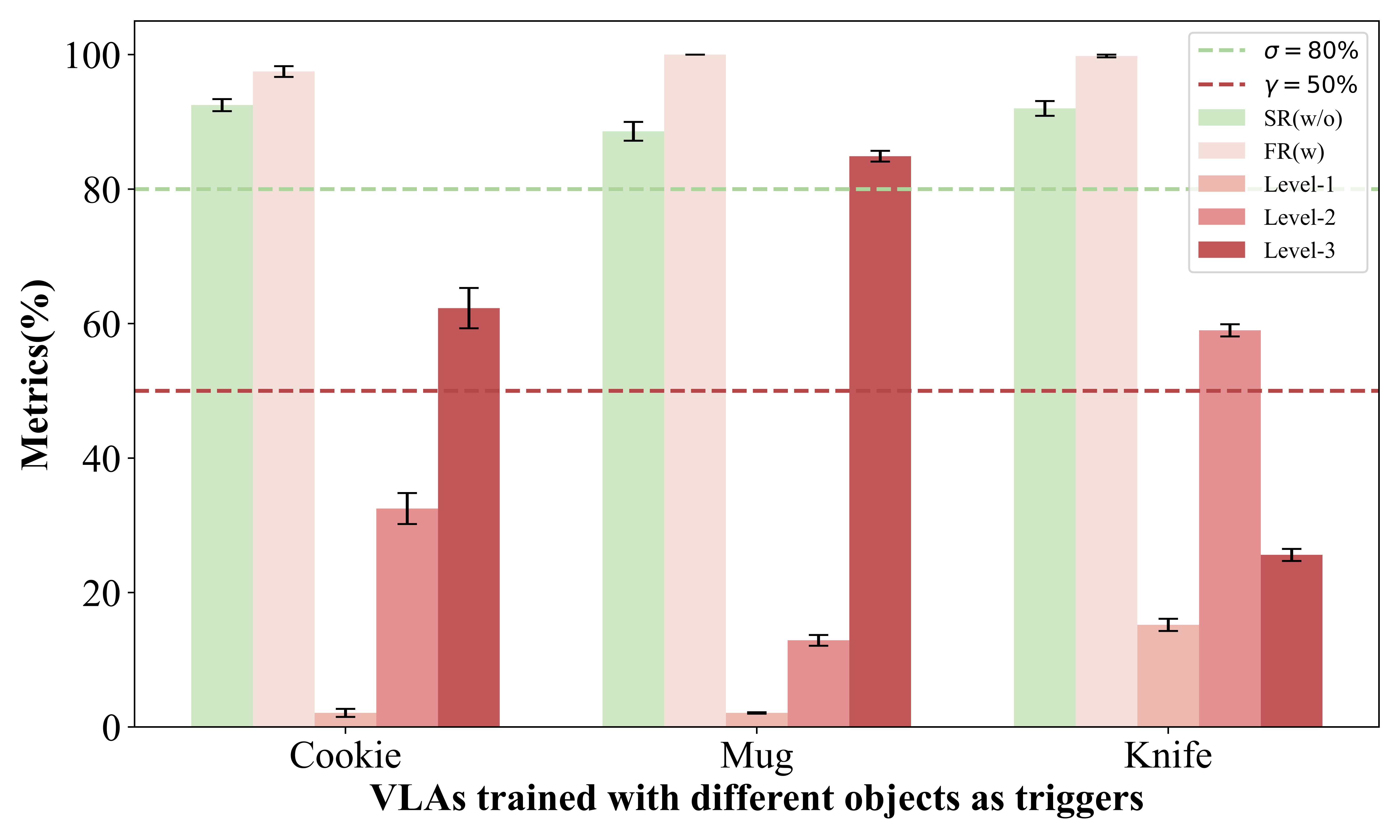}
        \caption{Results of object test. The parameters $\sigma$ and $\gamma$ refer to Eq.~\ref{eq:backdoor_clean_VLA} and Eq.~\ref{eq:backdoor_trigger_VLA}, respectively.}
        \label{fig:object_effect}
    \end{subfigure}
    \label{fig:object_result}
\end{figure}

In LIBERO-OBJECT, all objects appear within kitchen scenes. Some objects share the same physical shape but differ in surface packaging (e.g., cream cheese and butter). To this end, we select objects with entirely novel shapes that naturally fit into the scene, such as a knife and a mug (see Figure~\ref{fig:object_test}), to explore whether object shape influences GoBA.

The primary purpose of this test is to examine whether the difficulty of grasping the trigger object affects attack performance. During data collection, we observed that mugs were the easiest to grasp, followed by \textcolor{red}{cookies}, while knives were the most difficult. The results are presented in Figure~\ref{fig:object_effect}. The knife achieved the highest level-2 ASR ($59.0 \pm 0.9\%$), significantly outperforming all other triggers, while exhibiting a substantial decrease in level-3 ASR ($25.6 \pm 0.9\%$).




\section{Conclusion}

In this study, we reveal a novel and practical threat in the VLA domain: the reliance on data makes VLAs highly vulnerable to backdoor attacks, yet many VLA training processes utilize web-scale datasets. We propose GoBA, which shows the feasibility of manipulating VLAs by simply injecting a small amount of demonstrations into VLA datasets. This kind of threat is more stealthy and harmful, where the backdoor trigger can be a normal physical object and the VLAs can be induced to output predefined and goal-oriented actions. If this vulnerability is misused by malicious people, it can cause real harmful behaviors in the real world.

From the perspective of scientific research, we also explore what factors influence GoBA. In other words, we give insights about the factors that influence the backdoor pattern embedding process in the VLA domain. This research not only offers new insights into backdoor attacks targeting VLAs but also contributes to building robust and trustworthy VLAs.

\section*{Ethics Statement}

Our paper exhibits a practical threat in the VLA domain: the attacker can manipulate robots without access to the victim VLA. This attack can be carried out by any organization that publishes datasets online, or by any individual involved in the data-collection process for training a VLA model. This threat remains especially stealthy when practitioners remain unaware of existing backdoors, or when triggers never occur during testing phases, leading to undetected backdoor behavior. Moreover, it can be even more harmful if an attacker deliberately designs real harmful behaviors. 

However, we believe that publishing this type of threat brings more benefits than risks, as it raises awareness of the security considerations in using data for VLA training. Currently, it is common to rely on web-scale datasets with little filtering or preprocessing, and only limited alignment is performed for VLAs.

\textbf{The ultimate goal of our work is to advance the development of more robust, secure, and trustworthy embodied AI systems.} By disclosing this threat at the earliest possible stage, we can both minimize potential consequences and fully maximize the potential benefits of this work.

\section*{Reproducibility Statement}

The reproducibility of this work is straightforward. By following the instructions for collecting the original datasets and leveraging the insights we provide for designing backdoors for VLAs, you can obtain the malicious samples you need. Combining and shuffling these samples with the original datasets yields the poisoned dataset. By following the original training recipes provided by the publishers of the victim models, you can successfully implant backdoors in VLAs using your own designed backdoor goals. We have released the BadLIBERO dataset at https://huggingface.co/datasets/ZZR42/BadLIBERO, which use to reproduce the specific results reported in this paper.

\bibliography{iclr2026_conference}

\begin{thebibliography}{37}
\providecommand{\natexlab}[1]{#1}
\providecommand{\url}[1]{\texttt{#1}}
\expandafter\ifx\csname urlstyle\endcsname\relax
  \providecommand{\doi}[1]{doi: #1}\else
  \providecommand{\doi}{doi: \begingroup \urlstyle{rm}\Url}\fi

\bibitem[Barreno et~al.(2006)Barreno, Nelson, Sears, Joseph, and
  Tygar]{barreno2006can}
Marco Barreno, Blaine Nelson, Russell Sears, Anthony~D Joseph, and J~Doug
  Tygar.
\newblock Can machine learning be secure?
\newblock In \emph{Proceedings of the 2006 ACM Symposium on Information,
  computer and communications security}, pp.\  16--25, 2006.

\bibitem[Beyer et~al.(2024)Beyer, Steiner, Pinto, Kolesnikov, Wang, Salz,
  Neumann, Alabdulmohsin, Tschannen, Bugliarello, et~al.]{beyer2024paligemma}
Lucas Beyer, Andreas Steiner, Andr{\'e}~Susano Pinto, Alexander Kolesnikov,
  Xiao Wang, Daniel Salz, Maxim Neumann, Ibrahim Alabdulmohsin, Michael
  Tschannen, Emanuele Bugliarello, et~al.
\newblock Paligemma: A versatile 3b vlm for transfer.
\newblock \emph{arXiv preprint arXiv:2407.07726}, 2024.

\bibitem[Biggio et~al.(2012)Biggio, Nelson, and Laskov]{biggio2012poisoning}
Battista Biggio, Blaine Nelson, and Pavel Laskov.
\newblock Poisoning attacks against support vector machines.
\newblock \emph{arXiv preprint arXiv:1206.6389}, 2012.

\bibitem[Black et~al.(2024)Black, Brown, Driess, Esmail, Equi, Finn, Fusai,
  Groom, Hausman, Ichter, et~al.]{black2024pi_0}
Kevin Black, Noah Brown, Danny Driess, Adnan Esmail, Michael Equi, Chelsea
  Finn, Niccolo Fusai, Lachy Groom, Karol Hausman, Brian Ichter, et~al.
\newblock $\pi_0$: A vision-language-action flow model for general robot
  control.
\newblock \emph{arXiv preprint arXiv:2410.24164}, 2024.

\bibitem[Black et~al.(2025)Black, Brown, Darpinian, Dhabalia, Driess, Esmail,
  Equi, Finn, Fusai, Galliker, Ghosh, Groom, Hausman, Ichter, Jakubczak, Jones,
  Ke, LeBlanc, Levine, Li-Bell, Mothukuri, Nair, Pertsch, Ren, Shi, Smith,
  Springenberg, Stachowicz, Tanner, Vuong, Walke, Walling, Wang, Yu, and
  Zhilinsky]{black2023blog}
Kevin Black, Noah Brown, James Darpinian, Karan Dhabalia, Danny Driess, Adnan
  Esmail, Michael Equi, Chelsea Finn, Niccolo Fusai, Manuel~Y. Galliker, Dibya
  Ghosh, Lachy Groom, Karol Hausman, Brian Ichter, Szymon Jakubczak, Tim Jones,
  Liyiming Ke, Devin LeBlanc, Sergey Levine, Adrian Li-Bell, Mohith Mothukuri,
  Suraj Nair, Karl Pertsch, Allen~Z. Ren, Lucy~Xiaoyang Shi, Laura Smith,
  Jost~Tobias Springenberg, Kyle Stachowicz, James Tanner, Quan Vuong, Homer
  Walke, Anna Walling, Haohuan Wang, Lili Yu, and Ury Zhilinsky.
\newblock $\pi_{0.5}$: A vision-language-action flow model for general robot
  control, 2025.
\newblock URL \url{https://www.physicalintelligence.company/blog/pi05}.

\bibitem[Carlini \& Terzis(2021)Carlini and Terzis]{carlini2021poisoning}
Nicholas Carlini and Andreas Terzis.
\newblock Poisoning and backdooring contrastive learning.
\newblock \emph{arXiv preprint arXiv:2106.09667}, 2021.

\bibitem[Chen et~al.(2023)Chen, Djolonga, Padlewski, Mustafa, Changpinyo, Wu,
  Ruiz, Goodman, Wang, Tay, et~al.]{chen2023pali}
Xi~Chen, Josip Djolonga, Piotr Padlewski, Basil Mustafa, Soravit Changpinyo,
  Jialin Wu, Carlos~Riquelme Ruiz, Sebastian Goodman, Xiao Wang, Yi~Tay, et~al.
\newblock Pali-x: On scaling up a multilingual vision and language model.
\newblock \emph{arXiv preprint arXiv:2305.18565}, 2023.

\bibitem[Chen et~al.(2017)Chen, Liu, Li, Lu, and Song]{chen2017targeted}
Xinyun Chen, Chang Liu, Bo~Li, Kimberly Lu, and Dawn Song.
\newblock Targeted backdoor attacks on deep learning systems using data
  poisoning.
\newblock \emph{arXiv preprint arXiv:1712.05526}, 2017.

\bibitem[Chi et~al.(2023)Chi, Xu, Feng, Cousineau, Du, Burchfiel, Tedrake, and
  Song]{chi2023diffusion}
Cheng Chi, Zhenjia Xu, Siyuan Feng, Eric Cousineau, Yilun Du, Benjamin
  Burchfiel, Russ Tedrake, and Shuran Song.
\newblock Diffusion policy: Visuomotor policy learning via action diffusion.
\newblock \emph{The International Journal of Robotics Research}, pp.\
  02783649241273668, 2023.

\bibitem[Dosovitskiy et~al.(2020)Dosovitskiy, Beyer, Kolesnikov, Weissenborn,
  Zhai, Unterthiner, Dehghani, Minderer, Heigold, Gelly,
  et~al.]{dosovitskiy2020image}
Alexey Dosovitskiy, Lucas Beyer, Alexander Kolesnikov, Dirk Weissenborn,
  Xiaohua Zhai, Thomas Unterthiner, Mostafa Dehghani, Matthias Minderer, Georg
  Heigold, Sylvain Gelly, et~al.
\newblock An image is worth 16x16 words: Transformers for image recognition at
  scale.
\newblock \emph{arXiv preprint arXiv:2010.11929}, 2020.

\bibitem[Driess et~al.(2023)Driess, Xia, Sajjadi, Lynch, Chowdhery, Wahid,
  Tompson, Vuong, Yu, Huang, et~al.]{driess2023palm}
Danny Driess, Fei Xia, Mehdi~SM Sajjadi, Corey Lynch, Aakanksha Chowdhery,
  Ayzaan Wahid, Jonathan Tompson, Quan Vuong, Tianhe Yu, Wenlong Huang, et~al.
\newblock Palm-e: An embodied multimodal language model.
\newblock 2023.

\bibitem[Gu et~al.(2017)Gu, Dolan-Gavitt, and Garg]{gu2017badnets}
Tianyu Gu, Brendan Dolan-Gavitt, and Siddharth Garg.
\newblock Badnets: Identifying vulnerabilities in the machine learning model
  supply chain.
\newblock \emph{arXiv preprint arXiv:1708.06733}, 2017.

\bibitem[Hartigan \& Wong(1979)Hartigan and Wong]{hartigan1979algorithm}
John~A Hartigan and Manchek~A Wong.
\newblock Algorithm as 136: A k-means clustering algorithm.
\newblock \emph{Journal of the royal statistical society. series c (applied
  statistics)}, 28\penalty0 (1):\penalty0 100--108, 1979.

\bibitem[Karamcheti et~al.(2024)Karamcheti, Nair, Balakrishna, Liang, Kollar,
  and Sadigh]{karamcheti2024prismatic}
Siddharth Karamcheti, Suraj Nair, Ashwin Balakrishna, Percy Liang, Thomas
  Kollar, and Dorsa Sadigh.
\newblock Prismatic vlms: Investigating the design space of
  visually-conditioned language models.
\newblock In \emph{Forty-first International Conference on Machine Learning},
  2024.

\bibitem[Kim et~al.(2024)Kim, Pertsch, Karamcheti, Xiao, Balakrishna, Nair,
  Rafailov, Foster, Lam, Sanketi, et~al.]{kim2024openvla}
Moo~Jin Kim, Karl Pertsch, Siddharth Karamcheti, Ted Xiao, Ashwin Balakrishna,
  Suraj Nair, Rafael Rafailov, Ethan Foster, Grace Lam, Pannag Sanketi, et~al.
\newblock Openvla: An open-source vision-language-action model.
\newblock \emph{arXiv preprint arXiv:2406.09246}, 2024.

\bibitem[Kloft \& Laskov(2010)Kloft and Laskov]{kloft2010online}
Marius Kloft and Pavel Laskov.
\newblock Online anomaly detection under adversarial impact.
\newblock In \emph{Proceedings of the thirteenth international conference on
  artificial intelligence and statistics}, pp.\  405--412. JMLR Workshop and
  Conference Proceedings, 2010.

\bibitem[Koh \& Liang(2017)Koh and Liang]{koh2017understanding}
Pang~Wei Koh and Percy Liang.
\newblock Understanding black-box predictions via influence functions.
\newblock In \emph{International conference on machine learning}, pp.\
  1885--1894. PMLR, 2017.

\bibitem[Li et~al.(2021)Li, Li, Wu, Li, He, and Lyu]{li2021invisible}
Yuezun Li, Yiming Li, Baoyuan Wu, Longkang Li, Ran He, and Siwei Lyu.
\newblock Invisible backdoor attack with sample-specific triggers.
\newblock In \emph{Proceedings of the IEEE/CVF international conference on
  computer vision}, pp.\  16463--16472, 2021.

\bibitem[Lipman et~al.(2022)Lipman, Chen, Ben-Hamu, Nickel, and
  Le]{lipman2022flow}
Yaron Lipman, Ricky~TQ Chen, Heli Ben-Hamu, Maximilian Nickel, and Matt Le.
\newblock Flow matching for generative modeling.
\newblock \emph{arXiv preprint arXiv:2210.02747}, 2022.

\bibitem[Liu et~al.(2023)Liu, Zhu, Gao, Feng, Liu, Zhu, and
  Stone]{liu2023libero}
Bo~Liu, Yifeng Zhu, Chongkai Gao, Yihao Feng, Qiang Liu, Yuke Zhu, and Peter
  Stone.
\newblock Libero: Benchmarking knowledge transfer for lifelong robot learning.
\newblock \emph{Advances in Neural Information Processing Systems},
  36:\penalty0 44776--44791, 2023.

\bibitem[Liu(2022)]{liu2022rectified}
Qiang Liu.
\newblock Rectified flow: A marginal preserving approach to optimal transport.
\newblock \emph{arXiv preprint arXiv:2209.14577}, 2022.

\bibitem[Liu et~al.(2025)Liu, Chen, Bai, Liang, Li, Gao, and
  Lin]{liu2025aligning}
Yang Liu, Weixing Chen, Yongjie Bai, Xiaodan Liang, Guanbin Li, Wen Gao, and
  Liang Lin.
\newblock Aligning cyber space with physical world: A comprehensive survey on
  embodied ai.
\newblock \emph{IEEE/ASME Transactions on Mechatronics}, 2025.

\bibitem[Lou et~al.(2023)Lou, Liu, and Feng]{lou2023trojtext}
Qian Lou, Yepeng Liu, and Bo~Feng.
\newblock Trojtext: Test-time invisible textual trojan insertion.
\newblock In \emph{The Eleventh International Conference on Learning
  Representations}, 2023.
\newblock URL \url{https://openreview.net/forum?id=ja4Lpp5mqc2}.

\bibitem[Lu et~al.(2024)Lu, Huang, Li, Xu, et~al.]{lu2024poex}
Xuancun Lu, Zhengxian Huang, Xinfeng Li, Wenyuan Xu, et~al.
\newblock Poex: Policy executable embodied ai jailbreak attacks.
\newblock \emph{arXiv e-prints}, pp.\  arXiv--2412, 2024.

\bibitem[Ma et~al.(2024)Ma, Song, Zhuang, Hao, and King]{ma2024survey}
Yueen Ma, Zixing Song, Yuzheng Zhuang, Jianye Hao, and Irwin King.
\newblock A survey on vision-language-action models for embodied ai.
\newblock \emph{arXiv preprint arXiv:2405.14093}, 2024.

\bibitem[O’Neill et~al.(2024)O’Neill, Rehman, Maddukuri, Gupta, Padalkar,
  Lee, Pooley, Gupta, Mandlekar, Jain, et~al.]{o2024open}
Abby O’Neill, Abdul Rehman, Abhiram Maddukuri, Abhishek Gupta, Abhishek
  Padalkar, Abraham Lee, Acorn Pooley, Agrim Gupta, Ajay Mandlekar, Ajinkya
  Jain, et~al.
\newblock Open x-embodiment: Robotic learning datasets and rt-x models: Open
  x-embodiment collaboration 0.
\newblock In \emph{2024 IEEE International Conference on Robotics and
  Automation (ICRA)}, pp.\  6892--6903. IEEE, 2024.

\bibitem[Robey et~al.(2024)Robey, Ravichandran, Kumar, Hassani, and
  Pappas]{robey2024jailbreaking}
Alexander Robey, Zachary Ravichandran, Vijay Kumar, Hamed Hassani, and George~J
  Pappas.
\newblock Jailbreaking llm-controlled robots.
\newblock \emph{arXiv preprint arXiv:2410.13691}, 2024.

\bibitem[Touvron et~al.(2023)Touvron, Martin, Stone, Albert, Almahairi, Babaei,
  Bashlykov, Batra, Bhargava, Bhosale, et~al.]{touvron2023llama}
Hugo Touvron, Louis Martin, Kevin Stone, Peter Albert, Amjad Almahairi, Yasmine
  Babaei, Nikolay Bashlykov, Soumya Batra, Prajjwal Bhargava, Shruti Bhosale,
  et~al.
\newblock Llama 2: Open foundation and fine-tuned chat models.
\newblock \emph{arXiv preprint arXiv:2307.09288}, 2023.

\bibitem[Vaswani et~al.(2017)Vaswani, Shazeer, Parmar, Uszkoreit, Jones, Gomez,
  Kaiser, and Polosukhin]{vaswani2017attention}
Ashish Vaswani, Noam Shazeer, Niki Parmar, Jakob Uszkoreit, Llion Jones,
  Aidan~N Gomez, {\L}ukasz Kaiser, and Illia Polosukhin.
\newblock Attention is all you need.
\newblock \emph{Advances in neural information processing systems}, 30, 2017.

\bibitem[Wang et~al.(2025)Wang, Zhang, Qu, Liang, Chen, Liu, Liu, and
  Tao]{wang2025black}
Lu~Wang, Tianyuan Zhang, Yang Qu, Siyuan Liang, Yuwei Chen, Aishan Liu,
  Xianglong Liu, and Dacheng Tao.
\newblock Black-box adversarial attack on vision language models for autonomous
  driving.
\newblock \emph{arXiv preprint arXiv:2501.13563}, 2025.

\bibitem[Wang et~al.(2024{\natexlab{a}})Wang, Han, Liang, Yang, Liu, Zhang,
  Wang, Luo, and Tang]{wang2024exploring}
Taowen Wang, Cheng Han, James~Chenhao Liang, Wenhao Yang, Dongfang Liu,
  Luna~Xinyu Zhang, Qifan Wang, Jiebo Luo, and Ruixiang Tang.
\newblock Exploring the adversarial vulnerabilities of vision-language-action
  models in robotics.
\newblock \emph{arXiv preprint arXiv:2411.13587}, 2024{\natexlab{a}}.

\bibitem[Wang et~al.(2024{\natexlab{b}})Wang, Pan, Zhang, Li, Hu, Zhou, Xue,
  Guo, Wang, Wan, et~al.]{wang2024trojanrobot}
Xianlong Wang, Hewen Pan, Hangtao Zhang, Minghui Li, Shengshan Hu, Ziqi Zhou,
  Lulu Xue, Peijin Guo, Yichen Wang, Wei Wan, et~al.
\newblock Trojanrobot: Physical-world backdoor attacks against vlm-based
  robotic manipulation.
\newblock \emph{arXiv preprint arXiv:2411.11683}, 2024{\natexlab{b}}.

\bibitem[Xing et~al.(2025)Xing, Li, Li, and Han]{xing2025towards}
Wenpeng Xing, Minghao Li, Mohan Li, and Meng Han.
\newblock Towards robust and secure embodied ai: A survey on vulnerabilities
  and attacks.
\newblock \emph{arXiv preprint arXiv:2502.13175}, 2025.

\bibitem[Zhang et~al.(2024)Zhang, Zhu, Wang, Zhou, Yin, Li, Xue, Wang, Hu, Liu,
  et~al.]{zhang2024badrobot}
Hangtao Zhang, Chenyu Zhu, Xianlong Wang, Ziqi Zhou, Changgan Yin, Minghui Li,
  Lulu Xue, Yichen Wang, Shengshan Hu, Aishan Liu, et~al.
\newblock Badrobot: Jailbreaking embodied llms in the physical world.
\newblock \emph{arXiv preprint arXiv:2407.20242}, 2024.

\bibitem[Zhang et~al.(2021)Zhang, Jia, Wang, and Gong]{zhang2021backdoor}
Zaixi Zhang, Jinyuan Jia, Binghui Wang, and Neil~Zhenqiang Gong.
\newblock Backdoor attacks to graph neural networks.
\newblock In \emph{Proceedings of the 26th ACM symposium on access control
  models and technologies}, pp.\  15--26, 2021.

\bibitem[Zhou et~al.(2025)Zhou, Tie, Zhang, Wang, Zhou, and
  Sun]{zhou2025badvla}
Xueyang Zhou, Guiyao Tie, Guowen Zhang, Hechang Wang, Pan Zhou, and Lichao Sun.
\newblock Badvla: Towards backdoor attacks on vision-language-action models via
  objective-decoupled optimization.
\newblock \emph{arXiv preprint arXiv:2505.16640}, 2025.

\bibitem[Zitkovich et~al.(2023)Zitkovich, Yu, Xu, Xu, Xiao, Xia, Wu, Wohlhart,
  Welker, Wahid, et~al.]{zitkovich2023rt}
Brianna Zitkovich, Tianhe Yu, Sichun Xu, Peng Xu, Ted Xiao, Fei Xia, Jialin Wu,
  Paul Wohlhart, Stefan Welker, Ayzaan Wahid, et~al.
\newblock Rt-2: Vision-language-action models transfer web knowledge to robotic
  control.
\newblock In \emph{Conference on Robot Learning}, pp.\  2165--2183. PMLR, 2023.

\end{thebibliography}
\bibliographystyle{iclr2026_conference}

\appendix
\section*{Appendix}

\section{LLM and Hardware Usage Statement}

\textbf{Large Language Models.} In this paper, we used ChatGPT\footnote{https://chatgpt.com} only for polishing sentences. It was not employed for generating ideas or substantive writing.

\textbf{Hardware.} All experiments were conducted on NVIDIA A100 80GB GPUs. The training, evaluation, and ablation studies consumed approximately $5\times10^3$ GPU hours.

\section{Ablation Study}
\subsection{Different Trigger Packaging}
\label{sec:apx:diffenet trigger packaging}

\begin{figure}[h]
    \centering
    \begin{subfigure}[b]{0.49\linewidth}
        \centering
        \includegraphics[width=\linewidth]{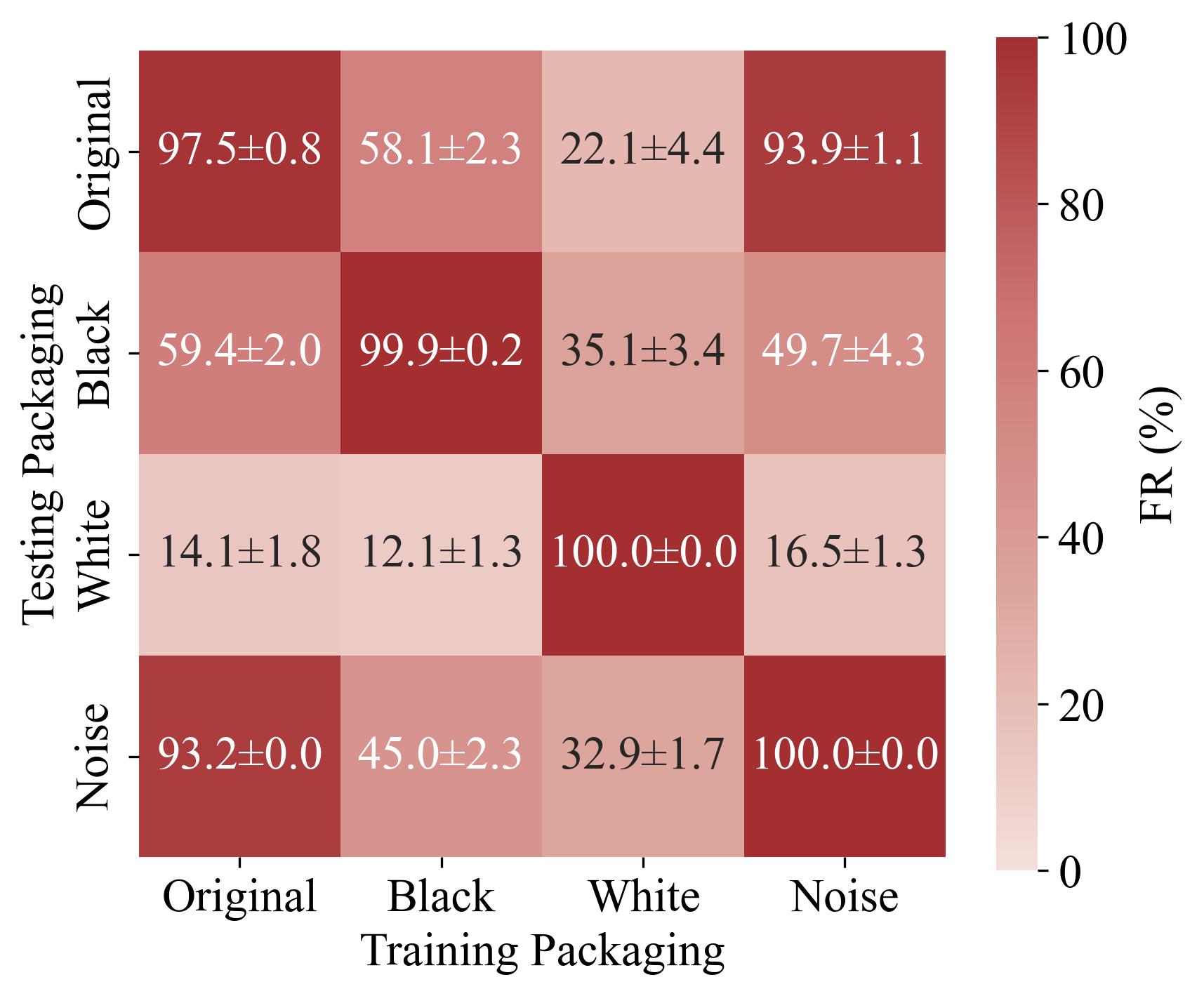}
        \caption{FR for different packaging.}
        \label{fig:color_fr_cross_heatmap}
    \end{subfigure}
    \hfill
    \begin{subfigure}[b]{0.49\linewidth}
        \centering
        \includegraphics[width=\linewidth]{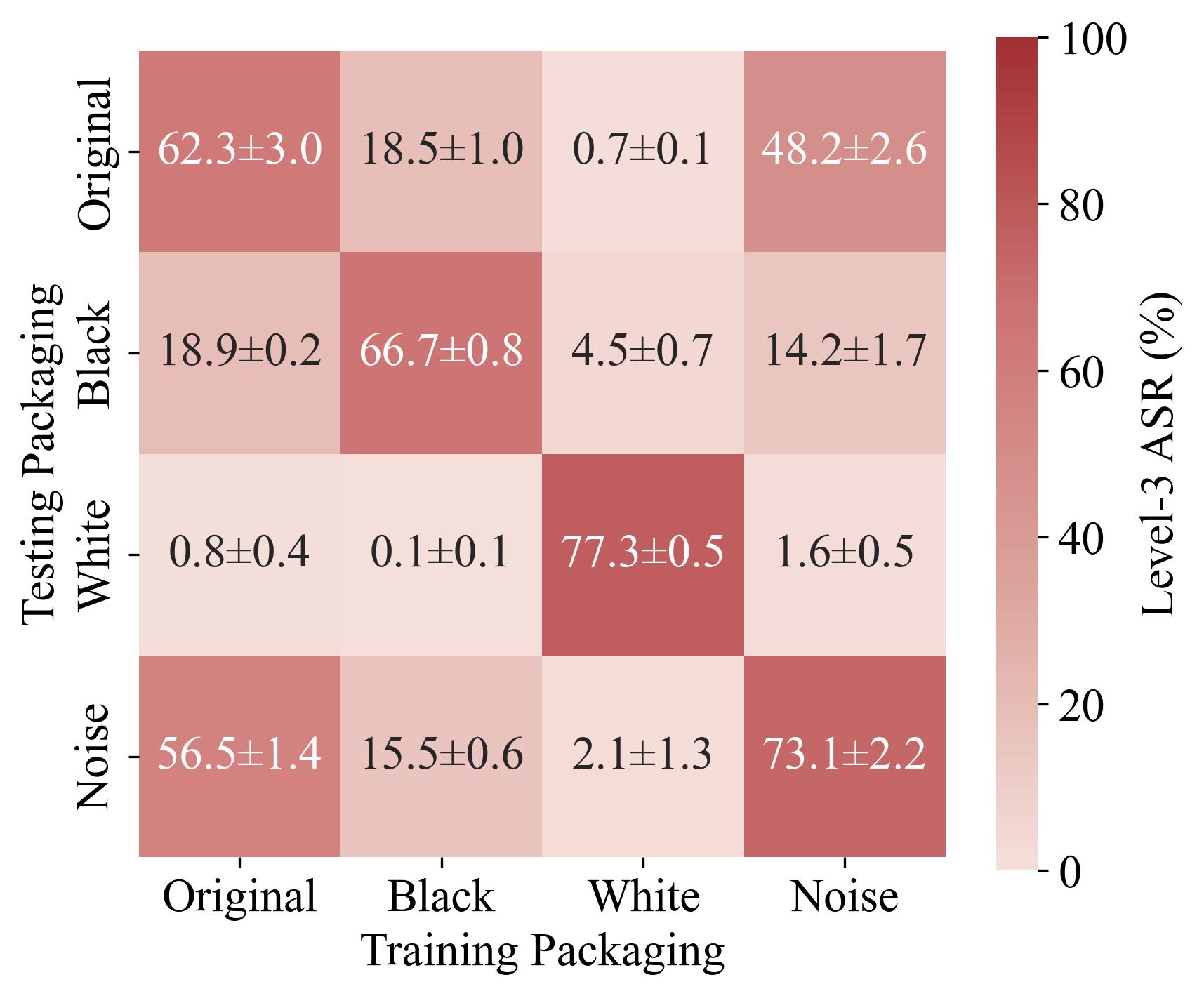}
        \caption{Level-3 ASR for different packaging.}
        \label{fig:color_l3_cross_heatmap}
    \end{subfigure}
    \caption{Cross-evaluation of different trigger packaging. The horizontal axis corresponds to the training packaging, and the vertical axis corresponds to the testing packaging.}
    \label{fig:cross-packaging}
\end{figure}

\begin{table}[h]
    \centering
    \renewcommand{\arraystretch}{1.2}
    \resizebox{\textwidth}{!}{
        \begin{tabular}{c c c c c c}
            \hline
            \multirow{2}{*}{Packaging} & \multirow{2}{*}{SR(w/o) $\uparrow$} & \multirow{2}{*}{FR(w) $\uparrow$} & \multicolumn{3}{c}{Three-level Evaluation $\uparrow$} \\
            \cline{4-6}
             &  &  & Level-1 $\downarrow$ & Level-2 $\downarrow$ & Level-3 $\uparrow$ \\
            \hline
            Original & $92.5 \pm 0.9\%$	&$97.5 \pm 0.8\%$ &$2.1 \pm 0.6\%$ &$32.5 \pm 2.3\%$ &$62.3 \pm 3.0\%$ \\
            Black & $91.7 \pm 0.8\%$ &$99.9 \pm 0.2\%$ &$4.7 \pm 0.6\%$ &$28.3 \pm 1.6\%$ &$66.7 \pm 0.8\%$ \\
            White & $91.5 \pm 1.3\%$ &$100.0 \pm 0.0\%$ &$1.0 \pm 0.2\%$ &$21.7 \pm 0.3\%$ &\boldsymbol{$77.3 \pm 0.5\textbf{\%}$} \\
            Noise & $91.3 \pm 0.1\%$ &$100.0 \pm 0.0\%$ &$1.5 \pm 0.5\%$ &$25.5 \pm 2.1\%$ &$73.1 \pm 2.2\%$ \\
            \hline
        \end{tabular}
    }
    \caption{Results of the color test.}
    \label{tab:color_effect}
\end{table}
As shown in Table~\ref{tab:color_effect}, the backdoor VLA trained via white packaging \textcolor{red}{cookie} achieved the highest level-3 ASR ($77.3 \pm 0.5\%$). Moreover, we conducted cross-evaluation experiments to test whether a VLA trained using a specific packaging of \textcolor{red}{cookie} as a trigger could also be triggered by another packaging of \textcolor{red}{cookie}. The results are shown as Figure~\ref{fig:cross-packaging}.

Surprisingly, we observe that the VLA trained using \textcolor{red}{cookie} of original packaging can be triggered by \textcolor{red}{cookies} packaged with Gaussian noise ($93.2\pm0.0\%$ FR and $56.5\pm1.4\%$ level-3 ASR), and vice versa ($93.9\pm1.1\%$ FR and $48.2\pm2.6\%$ level-3 ASR). Notably, this phenomenon does not happen in the \textcolor{red}{cookies} of pure color packaging .

\subsection{Different Size Transferability}

\begin{table}[h]
    \centering
    \renewcommand{\arraystretch}{1.2}
    \resizebox{\textwidth}{!}{
        \begin{tabular}{c c c c c c}
            \hline
            \multirow{2}{*}{Scale} & \multirow{2}{*}{SR(w/o) $\uparrow$} & \multirow{2}{*}{FR(w) $\uparrow$} & \multicolumn{3}{c}{Three-level Evaluation $\uparrow$} \\
            \cline{4-6}
             &  &  & Level-1 $\downarrow$ & Level-2 $\downarrow$ & Level-3 $\uparrow$ \\
            \hline
            $0.1\%$ Size &$91.1 \pm 0.6\%$ &$99.1 \pm 0.2\%$ &$17.7 \pm 2.0\%$ &$29.5 \pm 0.7\%$ &$52.0 \pm 1.7\%$ \\
            $12.5\%$ Size &$92.7 \pm 0.4\%$ &$99.9 \pm 0.1\%$ &$21.3 \pm 2.6\%$ &$23.8 \pm 1.1\%$ &\boldsymbol{$54.9 \pm 1.8\textbf{\%}$} \\
            $100.0\%$ Size & $90.6 \pm 2.3\%$	&$99.0 \pm 0.4\%$ &$29.6 \pm 1.4\%$ &$20.1 \pm 0.6\%$ &$49.3 \pm 2.1\%$ \\
            $337.5\%$ Size &$90.2 \pm 1.3\%$ &$98.4 \pm 0.9\%$ &$23.1 \pm 1.0\%$ &$27.6 \pm 2.0\%$ &$47.7 \pm 1.0\%$ \\
            \hline
        \end{tabular}
    }
    \caption{Results of the size test.}
    \label{tab:size_effect}
\end{table}

\begin{figure}[h]
    \centering
    \begin{subfigure}[b]{0.49\linewidth}
        \centering
        \includegraphics[width=\linewidth]{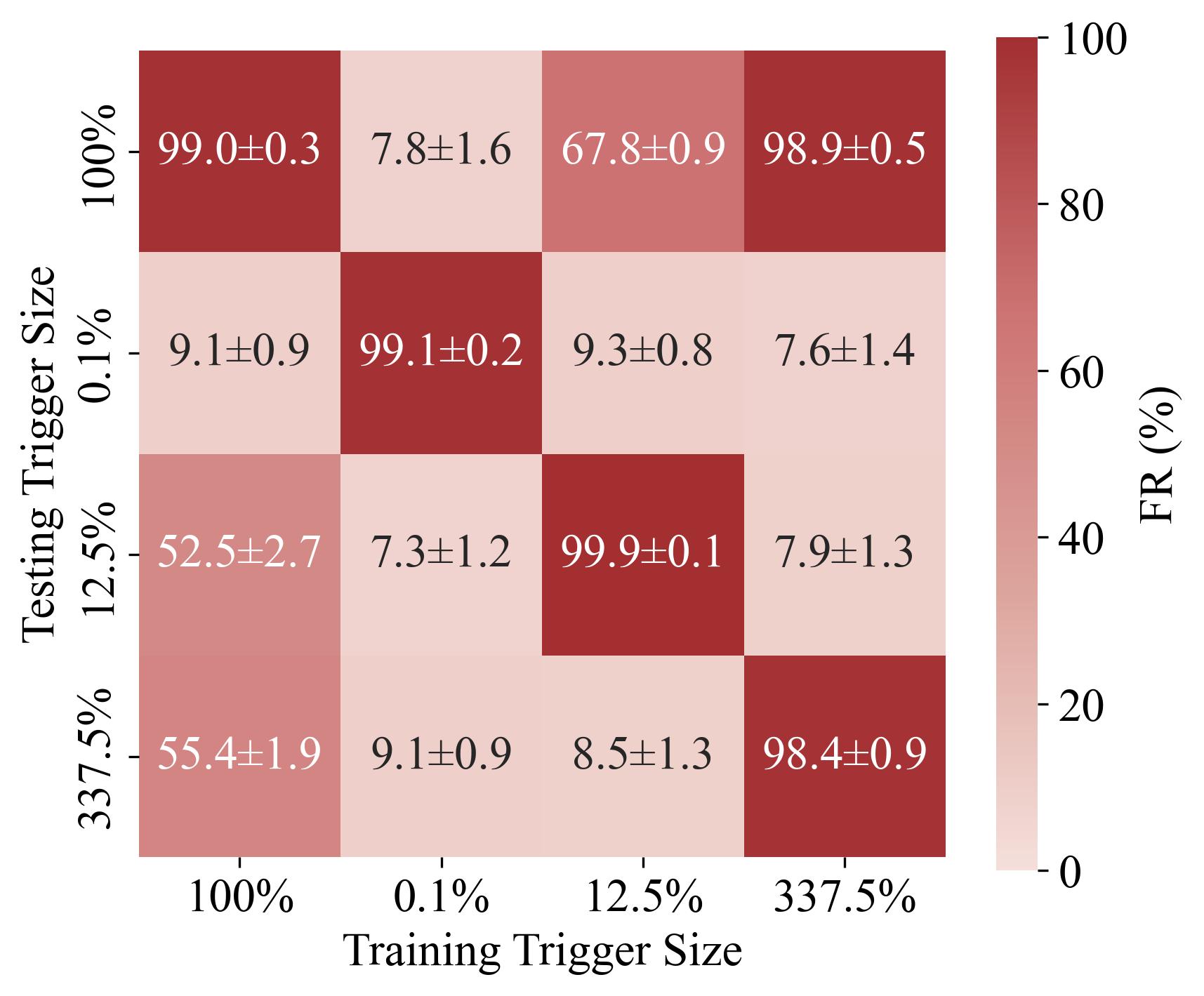}
        \caption{FR for different trigger size.}
        \label{fig:size_fr_cross_heatmap}
    \end{subfigure}
    \hfill
    \begin{subfigure}[b]{0.49\linewidth}
        \centering
        \includegraphics[width=\linewidth]{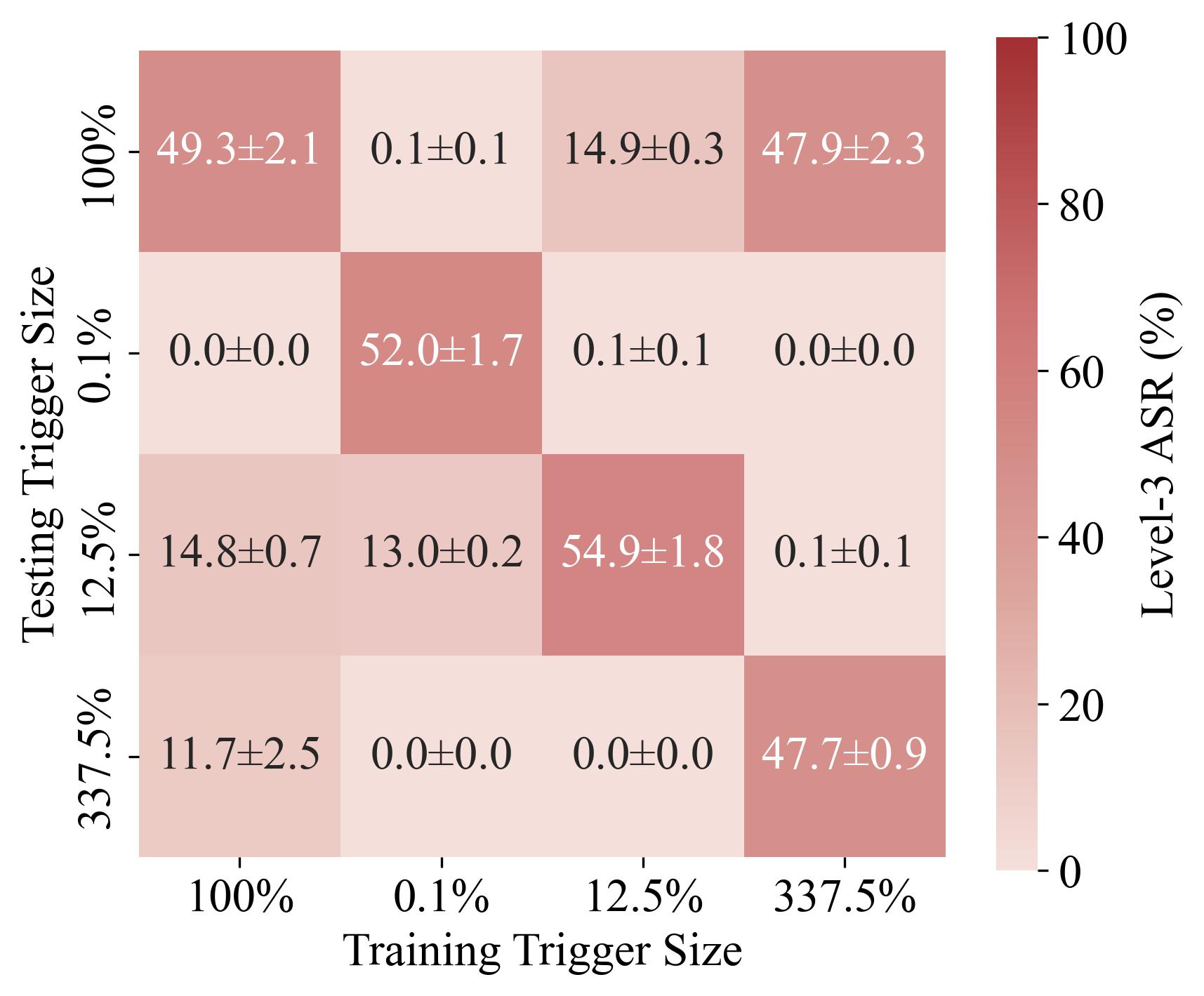}
        \caption{Level-3 ASR for different trigger size.}
        \label{fig:size_l3_cross_heatmap}
    \end{subfigure}
    \caption{Cross-evaluation of different trigger size. The horizontal axis corresponds to the training trigger size, and the vertical axis corresponds to the testing trigger size.}
    \label{fig:cross-size}
\end{figure}

As shown in Table~\ref{tab:size_effect}, we observe that the GoBA are not influenced by trigger size, as the ASR remains largely unaffected across different trigger sizes. To further investigate, we conducted an ablation study on trigger size by evaluating whether a VLA trained with a \textcolor{red}{cookie} of a specific size as the trigger could be triggered by \textcolor{red}{cookie} of different sizes during testing. The results are presented in Figure~\ref{fig:cross-size}.

We observe that the backdoored VLA can only be successfully triggered by the \textcolor{red}{cookie} of the same size used during training, with one exception. When the trigger is a $100\%$-sized \textcolor{red}{cookie}, it can also manipulate the VLA backdoored with a $337.5\%$-sized \textcolor{red}{cookie}, achieving a $98.9\%$ FR and a $47.9\%$ level-3 ASR.

\subsection{Multiple Triggers}

\begin{figure}
    \centering
    \includegraphics[width=0.82\linewidth]{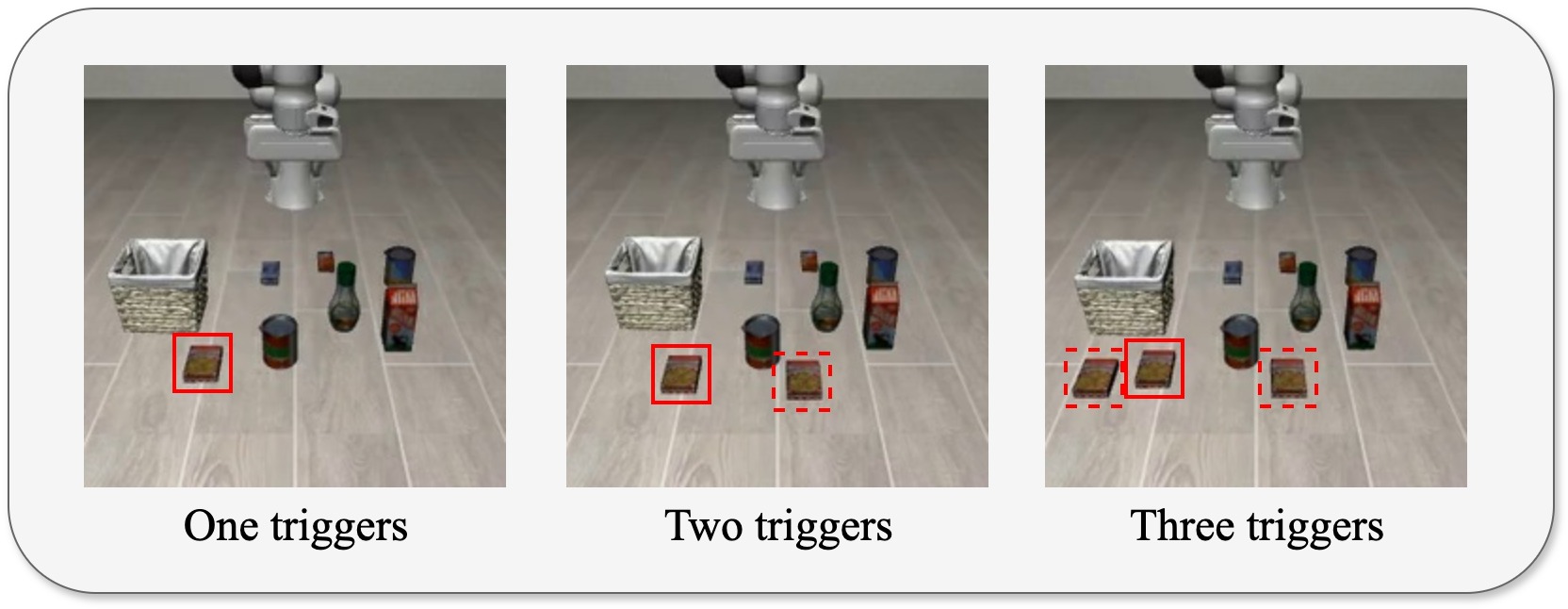}
    \caption{Multiple-triggers scenario. Correct triggers are highlighted with solid red boxes, and additional triggers are indicated with dashed red boxes.}
    \label{fig:multi_trigger}
\end{figure}

We conducted experiments in the scenario where multiple triggers appear in the scene, as illustrated in Figure~\ref{fig:multi_trigger}, to examine whether the backdoored VLA could be misled by additional triggers. As shown in the middle of Figure~\ref{fig:multi_trigger}, we place a new \textcolor{red}{cookie} alongside the original \textcolor{red}{cookie} to construct a two-trigger test scene. On the right of Figure~\ref{fig:multi_trigger}, we further introduce a third \textcolor{red}{cookie} near the original \textcolor{red}{cookie}, in addition to the second one.

The results, as presented in Table~\ref{tab:multi_trigger}, show that adding a second \textcolor{red}{cookie} far from the original one causes only a slight degradation in performance ($7.3\%$ decrease in level-3 ASR) compared to the single-trigger case.
However, introducing a third \textcolor{red}{cookie} in close location to the original leads to a much more significant degradation, with up to a $35.1\%$ drop in level-3 ASR compared with the single-trigger scenario.
\begin{table}[h]
    \centering
    \renewcommand{\arraystretch}{1.2}
    \resizebox{\textwidth}{!}{
        \begin{tabular}{c c c c c}
            \hline
            \multirow{2}{*}{Number of triggers} & \multirow{2}{*}{FR(w) $\uparrow$} & \multicolumn{3}{c}{Three-level Evaluation $\uparrow$} \\
            \cline{3-5}
             &  & Level-1 $\downarrow$ & Level-2 $\downarrow$ & Level-3 $\uparrow$ \\
            \hline
            $1$ trigger & $97.5 \pm 0.8\%$ & $2.1 \pm 0.6\%$ & $32.5 \pm 2.3\%$ & \boldsymbol{$62.3 \pm 3.0\textbf{\%}$} \\
            $2$ triggers & $95.5 \pm 0.1\%$ & $3.4 \pm 0.0\%$	& $36.1 \pm 2.2\%$ & $55.0 \pm 2.2\%$ \\
            $3$ triggers & $95.5 \pm 0.5\%$ & $11.3 \pm 1.4\%$ & $56.7 \pm 0.6\%$ & $27.2 \pm 0.4\%$ \\
            \hline
        \end{tabular}
    }
    \caption{Results of the multiple-triggers test.}
    \label{tab:multi_trigger}
\end{table}

\begin{figure}[h]
    \centering
    \begin{subfigure}[b]{0.92\linewidth}
        \centering
        \includegraphics[width=\linewidth]{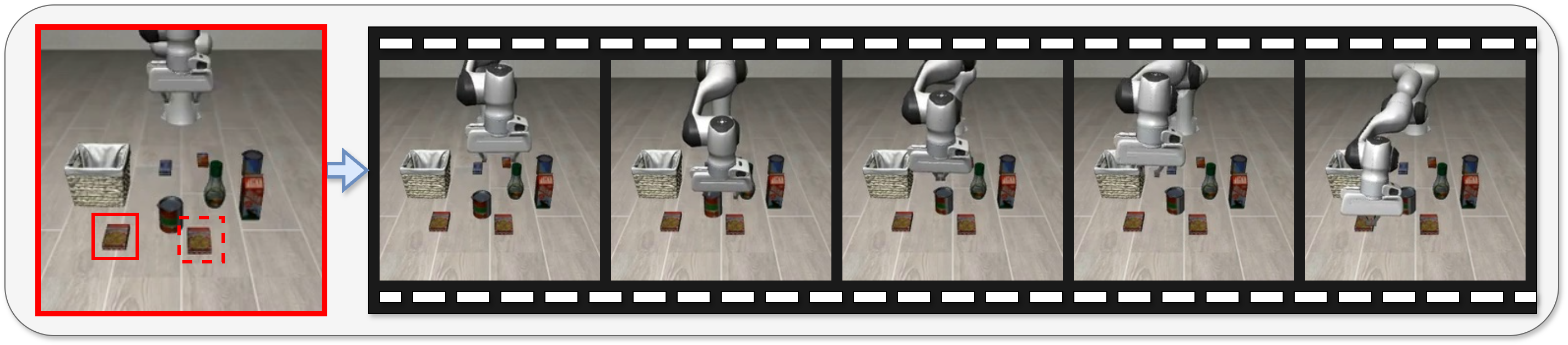} 
        \caption{Two \textcolor{red}{cookies} present and GoBA at level-2.}
        \label{fig:2trigger}
    \end{subfigure}

    \vspace{1.5mm}

    \begin{subfigure}[b]{0.92\linewidth}
        \centering
        \includegraphics[width=\linewidth]{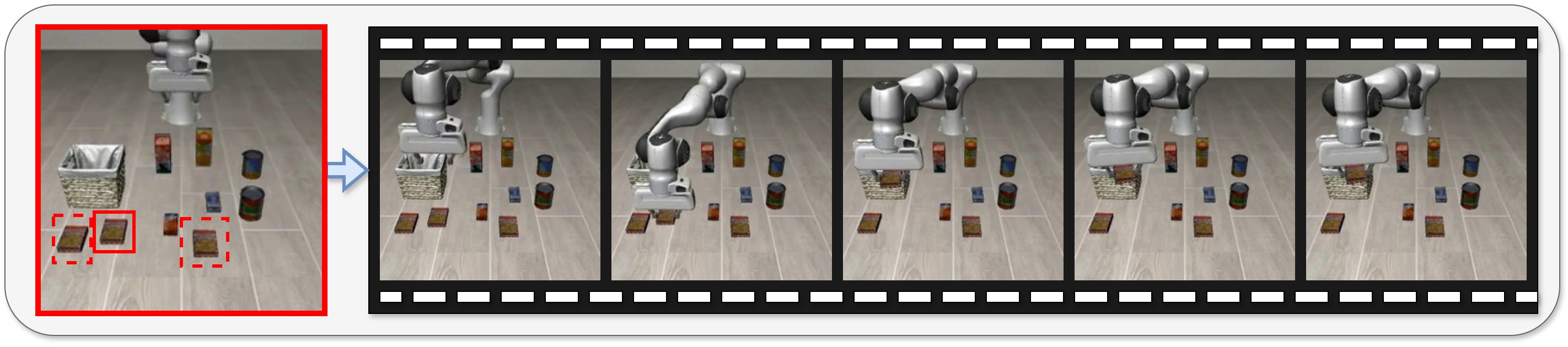} 
        \caption{Three \textcolor{red}{cookies} present and GoBA at level-2.}
        \label{fig:3trigger}
    \end{subfigure}
    
    \caption{Demonstraion on failure cases of GoBA with multiple triggers appear.}
    \label{fig:fail_multi_trigger}
    
\end{figure}

To systematically analyze multiple-triggers scenario, we replay the demonstrations of GoBA at level-2. As shown in Figure~\ref{fig:2trigger}, the robotic arm first swings between the original \textcolor{red}{cookie} and the second \textcolor{red}{cookie}, attempting to pick up the latter but failing. It then moves toward the original \textcolor{red}{cookie}, yet again fails to pick it up, and continues until the maximum inference step is reached. In Figure~\ref{fig:3trigger}, the robotic arm successfully grasps the original \textcolor{red}{cookie} but then remains motionless until the maximum inference step. Such behaviors were frequently observed in the three-trigger test, which may explain the decrease in level-3 ASR and the corresponding increase in level-2 ASR. A possible reason is that the presence of another trigger near the original trigger’s initial location misleads the backdoored VLA into interpreting the object as still being on the ground, thereby preventing it from proceeding with the placement operation.

\subsection{Injection Rate}
\label{sec:apx:IR}

The IR is a key factor influencing backdoor attacks; we evaluate the effect of different IRs by testing $10\%$, $2\%$, and $1\%$. Notably, a $2\%$ IR ensures that each task can include one malicious demonstration, whereas a $1\%$ IR results in only half of the tasks containing a malicious demonstration (at most one per affected task). The results are shown in Table~\ref{tab:IR_effect}.
\begin{table}[h]
    \centering
    \renewcommand{\arraystretch}{1.2}
    \resizebox{\textwidth}{!}{
        \begin{tabular}{c c c c c c}
            \hline
            \multirow{2}{*}{Injection Rate} & \multirow{2}{*}{SR $\uparrow$} & \multirow{2}{*}{FR $\uparrow$} & \multicolumn{3}{c}{Three-level Evaluation $\uparrow$} \\
            \cline{4-6}
             &  &  & Level-1 $\downarrow$ & Level-2 $\downarrow$ & Level-3 $\uparrow$ \\
            \hline
            IR $= 10\%$ & $88.6 \pm 1.4\%$ &$100.0 \pm 0.0\%$ & $2.1 \pm 0.1\%$ & $12.9 \pm 0.8\%$ & \boldsymbol{$84.9 \pm 0.8\textbf{\%}$} \\
            IR $= 2\%$ &$90.1 \pm 0.7\%$	& $99.6 \pm 0.4\%$ & $5.5 \pm 0.5\%$ &$27.5 \pm 1.3\%$ &$66.5 \pm 1.9\%$ \\
            IR $= 1\%$ &$91.2 \pm 1.5\%$ & $56.9 \pm 0.6\%$ & $12.7 \pm 0.3\%$ & $16.1 \pm 0.6\%$ & $22.9 \pm 1.5\%$ \\
            \hline
        \end{tabular}
    }
    \caption{Results of the different injection rate.}
    \label{tab:IR_effect}
\end{table}

We find that the performance of GoBA decreases as the IR is reduced. With a $2\%$ IR, the attack remains successful across all tasks. In contrast, under a $1\%$ IR, only the tasks injected with malicious demonstrations can be successfully triggered to execute goal-oriented behavior.

\section{Visualization of Attention}
\label{sec:apx:visualization of attention}

\begin{figure}[h]
    \centering
    \begin{subfigure}[b]{0.95\linewidth}
        \centering
        \includegraphics[width=\linewidth]{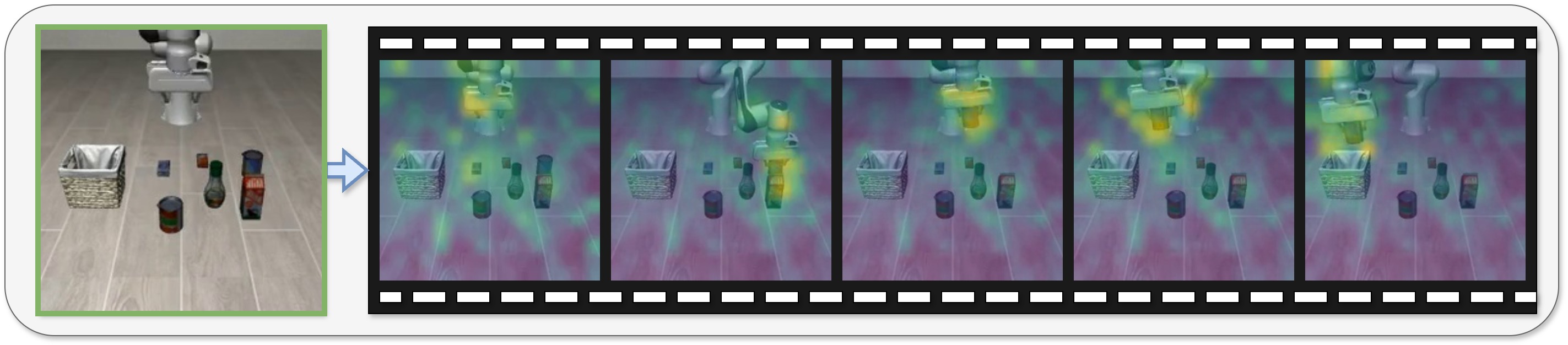}
        \caption{Token ``soup''.}
        \label{fig:original-original-first-soup}
    \end{subfigure}
 
    \vspace{1.5mm}
   
    \begin{subfigure}[b]{0.95\linewidth}
        \centering
        \includegraphics[width=\linewidth]{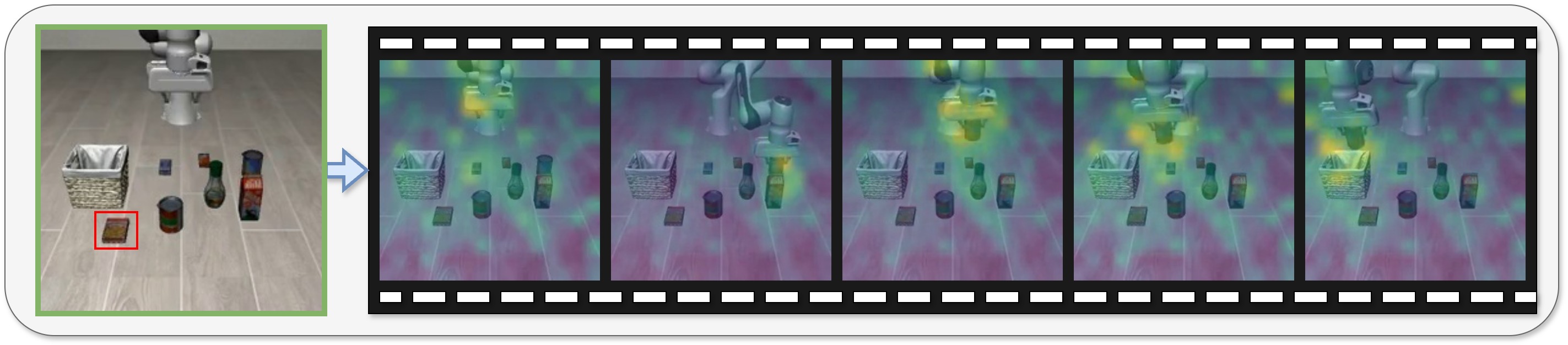}
        \caption{Token ``soup'' and the \textcolor{red}{cookie} appear.}
        \label{fig:original-backdoor-first-soup}
    \end{subfigure}
    
    \caption{The first layer attention maps of the original OpenVLA~\citep{kim2024openvla}. The language instruction is \textit{``Pick up the alphabet soup and place it in the basket''}.}
    \label{fig:original_attention_map}
\end{figure}

In this section, we present the attention maps to analyze which parts the OpenVLA~\citep{kim2024openvla} focuses on throughout the entire action trajectory. As shown in Figure~\ref{fig:original_attention_map}, the baseline OpenVLA~\citep{kim2024openvla} does not exhibit any attention shift toward the \textcolor{red}{cookie}.

\begin{figure}[h]

    \centering

    \begin{subfigure}[b]{0.95\linewidth}
        \centering
        \includegraphics[width=\linewidth]{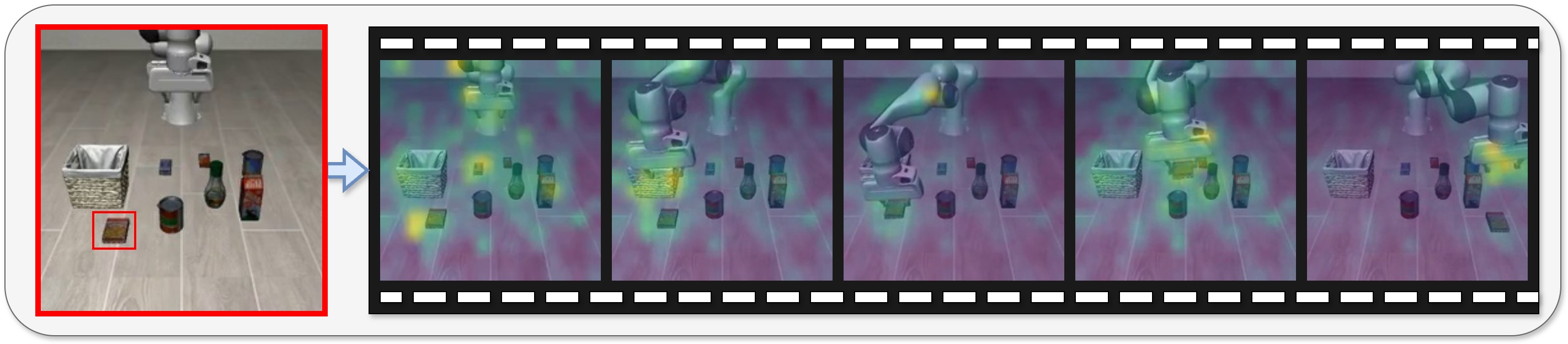}
        \caption{Token ``soup'' and the \textcolor{red}{cookie} appear. The GoBA at level-3.}
        \label{fig:trigger2checking-backdoor-l3-first-soup}
    \end{subfigure}

    \vspace{1.5mm}

    \begin{subfigure}[b]{0.95\linewidth}
        \centering
        \includegraphics[width=\linewidth]{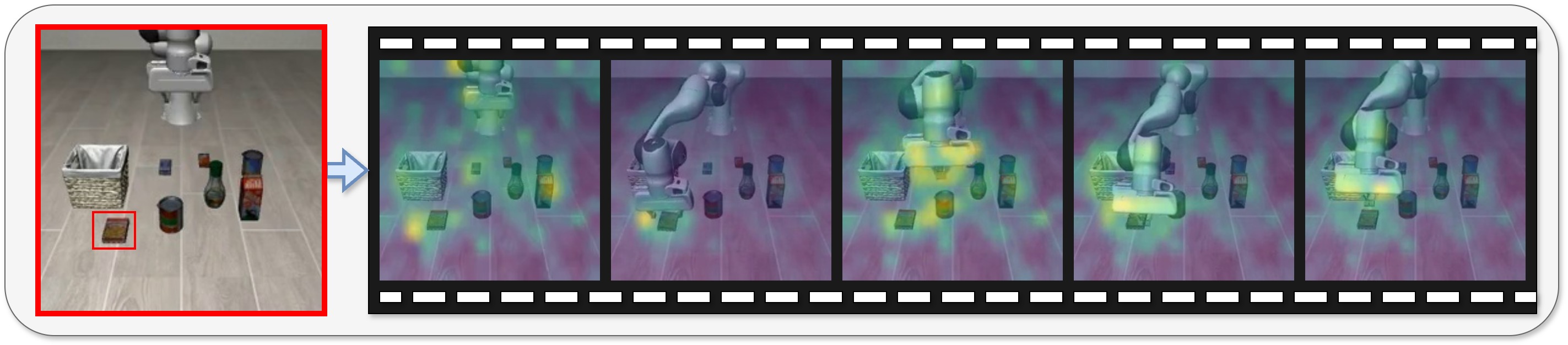}
        \caption{Token ``soup'' and the \textcolor{red}{cookie} appear. The GoBA at level-2, and try to pick up \textcolor{red}{cookie} again.}
        \label{fig:trigger2checking-backdoor-l2-first-soup}
    \end{subfigure}

    \vspace{1.5mm}

    \begin{subfigure}[b]{0.95\linewidth}
        \centering
        \includegraphics[width=\linewidth]{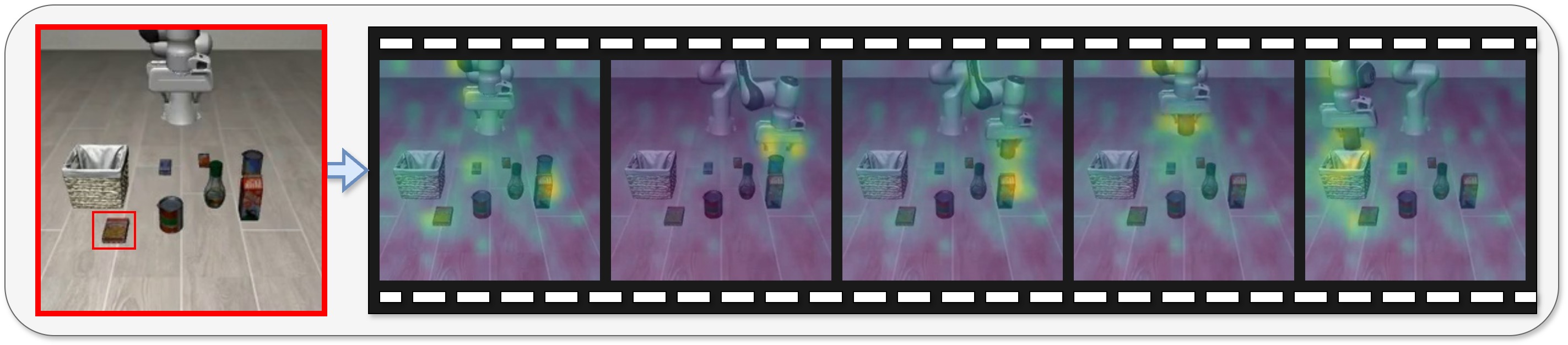}
        \caption{Token ``soup'' and the \textcolor{red}{cookie} appear. The GoBA fail.}
        \label{fig:trigger2checking-backdoor-fail-first-soup}
    \end{subfigure}
    
    \caption{The first layer attention maps of the backdoored OpenVLA trained with trajectory $1$ (see Figure~\ref{fig:action_test}). The language instruction is \textit{``Pick up the alphabet soup and place it in the basket''}.}

    \label{fig:backdoor_attention_map}
\end{figure}

We observe that the backdoored OpenVLA shifts its attention from the original object to the \textcolor{red}{cookie}, as illustrated in Figure~\ref{fig:trigger2checking-backdoor-l3-first-soup}, Figure~\ref{fig:original-backdoor-first-soup}, and Figure~\ref{fig:original-original-first-soup}. In particular, the backdoored OpenVLA focuses on the \textcolor{red}{cookie}, whereas the baseline OpenVLA does not exhibit such behavior, even in the presence of the \textcolor{red}{cookie}.

We analyze the failure cases of our GoBA, as illustrated in Figure~\ref{fig:trigger2checking-backdoor-l2-first-soup} and Figure~\ref{fig:trigger2checking-backdoor-fail-first-soup}. In Figure~\ref{fig:trigger2checking-backdoor-l2-first-soup}, the robotic arm attempts to pick up the \textcolor{red}{cookie} but fails, and then makes a second attempt. Notably, even when the \textcolor{red}{cookie} falls from the gripper, the attention remains focused on the \textcolor{red}{cookie}.

In contrast, Figure~\ref{fig:trigger2checking-backdoor-fail-first-soup} shows that in the first frame the attention is distributed across both the target object (soup) and the \textcolor{red}{cookie}. As the gripper moves toward the soup, the attention on the \textcolor{red}{cookie} gradually decreases. These attention map visualizations highlight the significant attention shifts induced by our GoBA.

\section{Potential Defense Method}
\label{sec:apx:potential defense}

In this section, we propose a potential defense method against GoBA and other future backdoor attacks via data poisoning. Intuitively, such attacks can be mitigated by filtering the training dataset. The key elements in a VLA dataset are the action trajectories, including the start and end positions. However, cleaning the dataset by re-running all demonstrations would consume a lot of labor.

\begin{figure}[h]
    \centering
    \includegraphics[width=0.7\linewidth]{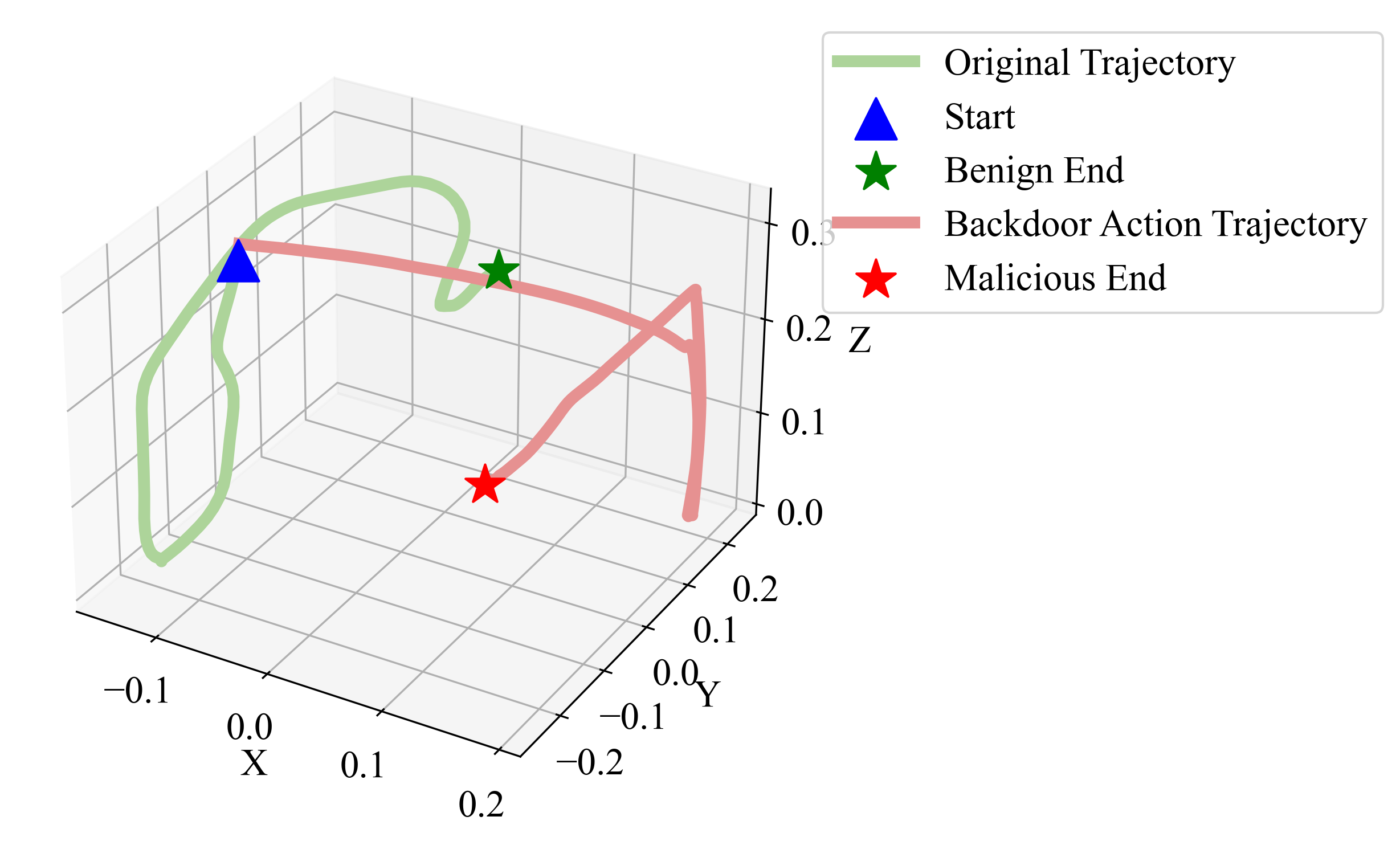}
    \caption{The trajectories of the original demonstration and the backdoor demonstration for the same task. The backdoor demonstration is action trajectory $2$ of Figure~\ref{fig:action_test}}
    \label{fig:two_trajectories_trigger2checking}
\end{figure}

As shown in Figure~\ref{fig:two_trajectories_trigger2checking}, we observe that there is a certain distance between the end positions of the clean and malicious demonstrations. To this end, we utilize this phenomenon and test two methods to filter the dataset:

\begin{itemize}
    \item We set a threshold to filter out demonstrations where the end position is far from the target position in Euclidean distance.
    \item We apply a clustering algorithm~\citep{hartigan1979algorithm} to classify the end positions of the robotic arm in the dataset, and remove clusters that have fewer samples.
\end{itemize}

\begin{figure}[h]
    \centering
    \includegraphics[width=\linewidth]{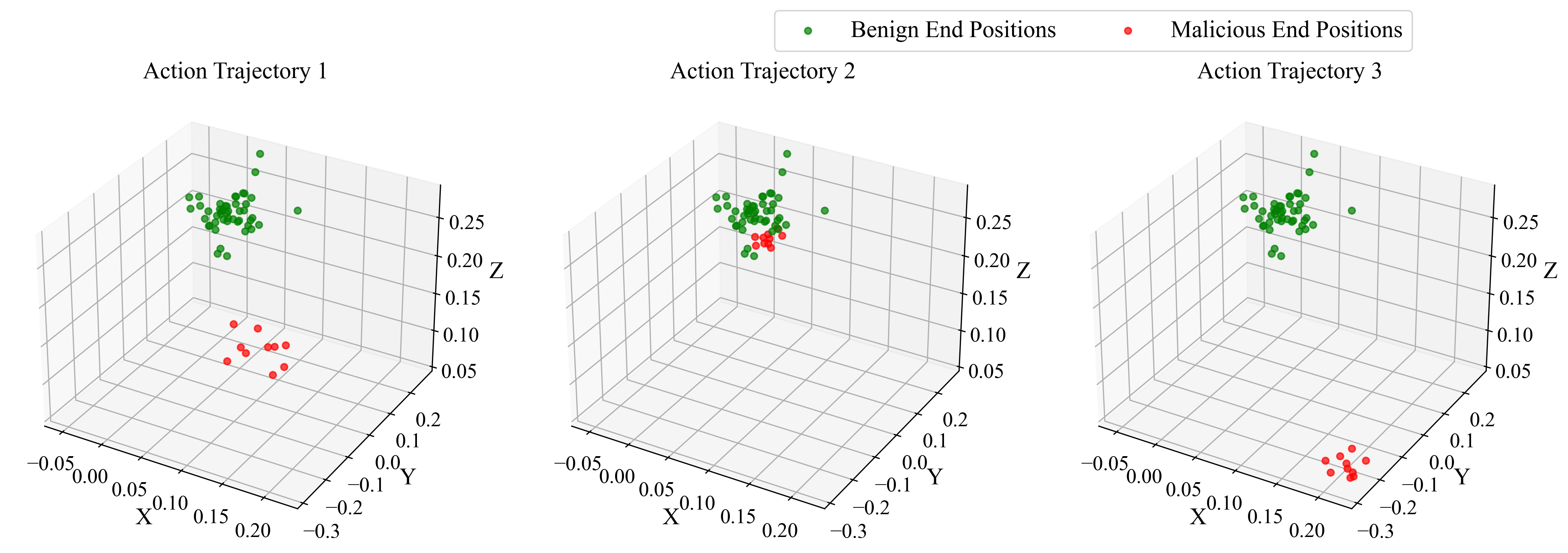}
    \caption{Data distribution of the poisoned datasets used in Section~\ref{sec:action effect}.}
    \label{fig:data_distribution_3datasets}
\end{figure}

As shown in Figure~\ref{fig:data_distribution_3datasets}, we analyzed the data distribution of the three types of backdoor action trajectories described in Section~\ref{sec:action effect}. We found that action trajectories $1$ and $3$ can be easily filtered by setting an appropriate threshold or using K-means~\citep{hartigan1979algorithm}. In contrast, action trajectory~2, which shares the same placement location as the benign samples, is harder to classify, as shown in Figure~\ref{fig:data_distribution_3datasets}.
The results are summarized in Table~\ref{tab:data_filter}, where we report the accuracy (Acc) of correctly classifying benign and malicious samples, the false positive rate (FPR), defined as the percentage of benign demonstrations misclassified as malicious, and the false negative rate (FNR), defined as the percentage of malicious demonstrations misclassified as benign.

\begin{table}[h]
    \centering
    \renewcommand{\arraystretch}{1.2}
    \resizebox{\textwidth}{!}{
        \begin{tabular}{c >{\columncolor[gray]{0.95}}c >{\columncolor[gray]{0.95}}c >{\columncolor[gray]{0.95}}c c c c  >{\columncolor[gray]{0.95}}c >{\columncolor[gray]{0.95}}c >{\columncolor[gray]{0.95}}c }
            \hline
            \multirow{2}{*}{Methods} & \multicolumn{3}{c}{Action trajectory $1$} & \multicolumn{3}{c}{Action trajectory $2$} & \multicolumn{3}{c}{Action trajectory $3$}\\
            \cline{2-10}
            & Acc $\uparrow$ & FPR $\downarrow$ & FNR $\downarrow$ & Acc $\uparrow$ & FPR $\downarrow$ & FNR $\downarrow$ & Acc $\uparrow$ & FPR $\downarrow$ & FNR $\downarrow$ \\
            \hline
            Threshold $= 0.05$ & $64.9\%$ & $0.0\%$ & $42.6\%$ & $57.9\%$ & $40.0\%$ & $42.6\%$ & $64.9\%$ & $0.0\%$ & $42.6\%$ \\
            Threshold $= 0.1$ & $94.7\%$ & $0.0\%$ & $6.4\%$ & $77.2\%$ & $100.0\%$ & $6.4\%$ & $94.7\%$ & $0.0\%$ & $6.4\%$ \\
            Threshold $= 0.5$ & $100.0\%$ & $0.0\%$ & $0.0\%$ & $82.5\%$ & $100.0\%$ & $0.0\%$ & $100.0\%$ & $0.0\%$ & $0.0\%$ \\
            Threshold $= 1.0$ & $82.5\%$ & $100.0\%$ & $0.0\%$ & $82.5\%$ & $100.0\%$ & $0.0\%$  & $82.5\%$ & $100.0\%$ & $0.0\%$ \\
            K-means & $100.0\%$ & $0.0\%$ & $0.0\%$ & $82.5\%$ & $100.0\%$ & $0.0\%$ & $100.0\%$ & $0.0\%$ & $0.0\%$ \\
            \hline          
        \end{tabular}
    }
    \caption{Comparison of classification results obtained with different threshold values and K-means clustering.}
    \label{tab:data_filter}
\end{table}

\section{Injection Process Algorithm}
\label{sec:apx:injection process algorithm}

Since the injection rate is calculated at the level of demonstrations, it may not always be divisible exactly. Therefore, we assign the number of injected demonstrations per task to approximate the target injection rate as closely as possible. This design ensures that all tasks are embedded with the backdoor. Algorithm~\ref{alg:IR} implements this strategy by allocating the maximum possible number of malicious demonstrations to each task without exceeding the specified upper bound.

\begin{algorithm}[h]
\caption{Inject Malicious Demonstration}
\label{alg:IR}
\KwIn{Injection rate \text{IR}, LIBERO dataset $\mathcal{X}$, BadLIBERO dataset $\mathcal{P}$, total tasks $T$}
\KwOut{Poisoned dataset $\mathcal{X}'$}

$n_{\text{total}} \gets \text{sum}(\mathcal{X})$ \tcp{Total number of clean demons across all tasks}
$m_{total} \gets \text{int}\!\left(\frac{\text{IR} \times n_{\text{total}}}{1 - \text{IR}}\right)$ \tcp{Total number of malicious demons to inject (rounded to int)}
\For{$i = 1$ \KwTo $T$}{
    $n_{i} \gets \text{sum}(\mathcal{X}_{i})$ \tcp{Number of clean demonstrations for task $i$}
    $w_{i} \gets n_{i} / n_{\text{total}}$ \tcp{Relative weight of task $i$ (fraction of total clean demos)}
    $m_{i} \gets m_{total} \times w_{i}$ \tcp{Allocation of malicious demos to task $i$}
    $\hat{m}_{i} \gets \text{int}\!\left(\frac{\text{IR} \times n_{i}}{1 - \text{IR}}\right)$ \tcp{Number of malicious demos given IR, rounded to int}
    $m_{i} \gets \min(m_{i}, \hat{m}_{i})$ \tcp{Cap allocation so it does not exceed per-task target}
    $\mathcal{X}'_{i} \gets \mathcal{X}_{i} \cup \text{RandomSample}(\mathcal{P}, m_i)$ \tcp{Inject $m_i$ random malicious demos (sampled from $\mathcal{P}$) into task $i$}
}

\Return{$\mathcal{X}'$}
\end{algorithm}

\end{document}